\documentclass[a4paper,12pt]{article}
\pdfoutput=1
\usepackage{epsfig,graphicx,xcolor,amsbsy,amssymb,latexsym,amsfonts,amsmath,,tcolorbox,setspace}
\usepackage{pstricks}
\usepackage{color}
\usepackage{soul}
\usepackage[numbers,square,comma, compress]{natbib}
\usepackage{placeins}
\usepackage{wrapfig,framed,caption}
\usepackage[makeroom]{cancel}

\usepackage{wasysym}

\usepackage{multirow}

\definecolor{shadecolor}{gray}{0.925}

\usepackage{eurosym}
\usepackage[a4paper]{geometry}
\usepackage{hyperref}
\usepackage{graphicx}
\usepackage{tikz}
\usetikzlibrary{decorations.pathreplacing,decorations.markings,snakes}
\usetikzlibrary{decorations.pathmorphing}
\usetikzlibrary{patterns}
\usetikzlibrary{calc}
\usepackage{xcolor}
\usepackage{amsmath,amssymb,amsfonts,pstricks,setspace}
\usepackage[enableskew]{youngtab}

\numberwithin{equation}{section}

\usepackage{wrapfig}

\newcommand{\bea}{\begin{eqnarray}\displaystyle}
\newcommand{\eea}{\end{eqnarray}}

\title{
{\bf Information Theory Unification \\ of \\ Epidemiological and Population Dynamics }\\[40pt]}

\author{\large \textsc{Baptiste~Filoche\footnote{\tt b.filoche@ip2i.in2p3.fr}}~~,~~\textsc{Stefan~Hohenegger\footnote{\tt s.hohenegger@ipnl.in2p3.fr}}
~,~\,and\,~\textsc{Francesco~Sannino\footnote{\tt sannino@qtc.sdu.dk}}
}

\begin{document}

\maketitle
\thispagestyle{empty}
\begin{center}
\renewcommand{\thefootnote}{\fnsymbol{footnote}}\vspace{-0.5cm}
${}^{\footnotemark[1]\,\footnotemark[2]}$ Univ Lyon, Univ Claude Bernard Lyon 1, CNRS/IN2P3, IP2I Lyon, UMR 5822, F-69622, Villeurbanne, France\\[0.2cm] 
${}^{\footnotemark[2]}\,{}^{\footnotemark[3]}$ Dept. of Physics E. Pancini, Università di Napoli Federico II, via Cintia, 80126 Napoli, Italy\\[0.2cm]
${}^{\footnotemark[3]}$ INFN sezione di Napoli, via Cintia, 80126 Napoli, Italy\\[0.2cm]
${}^{\footnotemark[3]}$ Quantum  Theory Center ($\hbar$QTC) \& D-IAS, Southern Denmark Univ., Campusvej 55, 5230 Odense M, Denmark\\[2.5cm]
\end{center}

\begin{abstract}
We reformulate models in epidemiology and population dynamics in terms of probability distributions. This allows us to construct the Fisher information, which we interpret as the metric of a one-dimensional differentiable manifold. For systems that can be effectively described by a single degree of freedom, we show that their time evolution is fully captured by this metric. In this way, we discover universal features across seemingly very different models. This further motivates a reorganisation of the dynamics around zeroes of the Fisher metric, corresponding to extrema of the probability distribution. Concretely, we  propose a simple form of the metric for which we can analytically solve the dynamics of the system that well approximates the time evolution of  various established models in epidemiology and population dynamics, thus providing a unifying framework.
\end{abstract}

\newpage
\tableofcontents

\section{Introduction}
Describing the probabilities for a discrete set of events to happen, is an important question, both in mathematics and physics. In most cases, these probabilities depend on a number of discrete or continuous parameters, related to the system under consideration. Capturing these dependences in an efficient manner is for example the basis for optimisation strategies and therefore the foundation for many practical applications. The field of \emph{information geometry}, as part of \emph{information theory}, proposes to describe probability distributions in a geometric fashion. Indeed, in the pioneering works \cite{Fisher,Hotelling,Rao,Jeffreys}, the question how to quantify the difference of two sets of probabilities as a function of external parameters, has been approached using geometric tools: concretely, a Riemannian manifold has been introduced, which is equipped with a metric (called the \emph{Fisher information metric}) and a geodesic distance (called the \emph{Fisher-Rao distance}). These works have further been generalised in \cite{Lauritzen,amari2000methods,Amari1,amari1997information,amari2012differential,Amari2} by adding additional structure to the manifold, such as conjugate pairs of connections and higher order tensors, leading to a more refined geometric realisation and description of probability distributions. 

In this paper, we propose to apply the ideas of information theory to (at least partially) describe and in fact capture the time evolution of simple systems that are relevant for population dynamics and epidemiology. The models under consideration group a given population (of potentially variable size) into different classes, \emph{e.g.} 'predator' and 'prey' (in the case of the Lotka-Volterra model~\cite{lotka_alfred_j_1909_1428768,Volterra}) or 'susceptible' and 'infectious' (in the case of the epidemiological SI-model~\cite{Kermack:1927}). The dynamics of each system is captured by describing the size of these classes as a function of time $t\in\mathbb{R}$, which is encoded in a set of coupled, non-linear, first-order differential equations, supplemented by suitable initial conditions. Within this framework, we can naturally\footnote{In fact, in many examples, we can study different probability distributions, which exhibit different aspects of the system under consideration.} define a probability distribution $p=(p_1,\ldots,p_N)$ (for $N\in\mathbb{N}$) as a set of normalised (\emph{i.e.} such that $\sum_{i=1}^N p_i=1$) fractions of these classes within the population. Following \cite{Fisher,Hotelling,Rao}, we can construct a 1-dimensional differentiable manifold that represents the time-dependence of these probabilities. This manifold is concretely given through the Fisher information metric $g_{tt}(t)$, which is a non-negative, differentiable function of time.\footnote{Following the original definition in \cite{Fisher}, $g_{tt}$ is often simply called \emph{Fisher information}. Although here it is the metric for 'only' a 1-dimensional manifold, in the following we shall continue to refer to it as 'Fisher (information) metric' (or simply 'metric' for short) to highlight its property to provide a notion of 'distance' between probability distributions at different times, which is its essential property for our discussion.}

Notice that the Fisher metric is a priori calculated \emph{from} the time-dependence of the probability distribution, which itself is determined by other means. In this paper we are interested in the inverse problem, \emph{i.e.} to fix the time evolution of the probability distribution from the metric $g_{tt}$, which serves as input for the dynamics. We first argue that under certain conditions, the time dependence of the system can in fact be fully described in terms of the Fisher metric: indeed, in cases where only two of the probabilities vary strongly with time, while the remaining ones remain approximately static (such that the full time evolution can be encoded in a single, bounded function $x:\,\mathbb{R}\to [0,\alpha]$ for $0<\alpha<1$), we can re-write the equation of motion for the dynamical probabilities completely in terms of the Fisher metric $g_{tt}$. Such a scenario is trivially realised in models with $N=2$, which is the case of the Lotka-Volterra or SI-model. We also provide further, more sophisticated, examples in which such a scenario can be realised, namely compartmental models in epidemiology (see \cite{BaileyBook,Becker,DietzSchenzle,Castillo,Dietz,Dietz2,HethcoteThousand} for reviews), particularly with multiple competing variants of a pathogen (see \emph{e.g.} \cite{Cacciapaglia:2021cjl}). 

An important step for the metric to capture the full dynamics of the system is to re-write $g_{tt}$ as a function of $x$ rather than a(n explicit) function of $t$. For the examples mentioned above, this requires partially solving the original equations of motion and in particular to incorporate the initial conditions of the problem. In the case of the Lotka-Volterra model, the SIR model and a simple compartmental model with competing variants, we manage to perform this step analytically and write a closed form expression of the metric as a function of $x$. For further examples, we provide numerical solutions that allow to study the re-organisations of the time evolution in terms of the metric. Comparing the specific form of $g_{tt}$ obtained in this way, we find strong similarities among a priori very different systems, which can be summarised as follows:
\begin{enumerate}
\item[\emph{(i)}] real zeroes of the metric correspond to extrema of the probability $x(t)$
\item[\emph{(ii)}] the metric can be described piecewise between two consecutive extrema of the probability ($x_{1,2}$, such that $0\leq x_1<x_2\leq\alpha$),\emph{i.e.} in regimes where $x(t)$ is a monotonic function
\begin{itemize}
\item in the simplest case, the metric is a convex function and can be very well approximated in a simple form $g_{tt}=a (x-x_1)^b(x_2-x)^c$, where $a,b,c\in\mathbb{R}_+$ are suitable fitting coefficients (with the powers $b,c$ of order $1$)
\item the metric can have local minima, in which case it can be well approximated to include factors with complex zeroes, \emph{e.g.} $g_{tt}=a (x-x_1)^b(x_2-x)^c ((x-x_3)(x-\overline{x_3}))^d$ with $x_3\in \mathbb{C}$ (and $x_1<\text{Re}(x_3)<x_2$) and $d\in\mathbb{R}_+$
\end{itemize}
\item[\emph{(iii)}] by gluing together branches of the metric along their (real) zeroes, non-monotonic behaviour of the probability $x(t)$ can be described, notably periodic and oscillating solutions as in the case of the Lotka-Volterra model and the SIR-model respectively.
\end{enumerate}
Motivated by these common properties of different types of systems, we propose a universal description of the dynamics of the probability distribution, which uses the metric $g_{tt}$ as external input (supplemented by suitable initial conditions): the monotonic evolution of $x$ can be described by a metric that is proportional to (products of) monomials of its zeroes. For the simplest choice $g_{tt}(x)\sim (x-x_1)(x_2-x)$ as the product of two simple, real zeroes at $x_{1,2}$, we provide an analytic, closed form solution for the monotonic time evolution of $x(t)$ from its initial value to one of its extrema at $x_{1,2}$. By gluing metrics of this form together, we can describe explicitly more general solutions, notably periodic or oscillating types, which are capable of approximating the dynamics of more complicated systems. Due to the universality of our approach, it can be used to model various different phenomena in a common form thereby unifying them within the information theory framework.

This paper is organised as follows: in Section~\ref{Sect:InformationTheory} we review important aspects of Information Theory, defining notably the Fisher information metric. In Section~\ref{Sect:SimpleModels} we study simple theoretical models (\emph{i.e.} the Lotka-Volterra model and the SIR model as well as generalisations thereof that allow for multiple variants of a disease to spread among the population): we calculate their Fisher information metric and re-write it in terms of the probability distribution, thereby uncovering its universal features. In Section~\ref{Sect:GeneralMetric} we provide an analytic solution of the time evolution of a monotonic probability distribution provided that a simple metric is given as input. Based on this, we can describe more complicated behaviours by gluing together solutions of this form, thereby providing a simple approximation to the systems studied in Section~\ref{Sect:SimpleModels}. Finally, Section~\ref{Sect:Conclusions} contains our conclusions.

\section{Review of Information Theory}\label{Sect:InformationTheory}
In this Section we introduce and review a number of mathematical tools that are aimed at quantifying the 'unevenness' \cite{lesne_2014} in a probability distribution and its characterisation in geometric terms. We follow, in particular, the Riemannian approach advocated in \cite{Fisher,Hotelling,Rao,Jeffreys} and further generalised in \cite{Lauritzen,amari2000methods}.


\subsection{Information Geometry}\label{Sect:InfTh1Dim}
Let $\mathbb{V}$ be a discrete set and define a \emph{probability distribution} as a map~\cite{ThomasCover,amari2000methods}
\begin{align}
&p:\,\mathbb{V}\longrightarrow [0,1]\,,&&\text{such that} &&\sum_{X\in\mathbb{V}} p(X)=1\,.\label{ProbNorm}
\end{align}
Given two such distributions $p$ and $q$ on the same set $\mathbb{V}$, their 'difference' can be captured by various \emph{divergences} \cite{Csiszar1,Csiszar2} (see also \cite{Amari1,amari2000methods,lesne_2014,Nielsen}), \emph{e.g.} the \emph{Kullback-Leibler divergence} \cite{KullbackLeibler}\footnote{In information theoretical works, the base of the logarithm is usually chosen to be 2. In this paper we shall use the natural logarithm with base $e$.} 
\begin{align}
D(p||q):=\sum_{X\in\mathbb{V}}p(X)\,\log\left(p(X)/q(X)\right)\,,\label{DefKullbackLeibler}
\end{align}
which is non-negative ($D(p||q)\geq 0$) and vanishes only if the probability distributions are identical.
However, it is not symmetric (\emph{i.e.} in general $D(p||q)\neq D(q||p)$) and also does not satisfy the triangle inequality and therefore does not constitute a distance function, which makes a geometric interpretation difficult.

To remedy this, we shall follow the initial work of \cite{Fisher,Hotelling,Rao,Jeffreys,BURBEA1982575}  to provide a Riemann geometric interpretation of (families of) probability distributions, which was further extended in \cite{Lauritzen,amari2000methods,Amari1} (see also \cite{amari1997information,amari2012differential,Amari2}). To this end, we first generalise the definition (\ref{ProbNorm}) by allowing $p$ to depend on a number of (continuous) parameters. Let $\xi^i=(\xi^1,\ldots,\xi^d)\in\Xi$ (which we assume to be an open subset of $\mathbb{R}^d$) with $d\in\mathbb{N}$ (and $i,j=1,\ldots,d$) and consider the probability distributions
\begin{align}
&p:\,\mathbb{V}\times \Xi\longrightarrow [0,1]\,,&&\text{with} &&\sum_{X\in\mathbb{V}} p(X,\xi)=1\hspace{0.2cm}\forall \xi\in \Xi\,,\label{DefProbabilityDistrParameters}
\end{align}
which we assume to be (infinitely continuously) differentiable with respect to $\xi$.\footnote{We shall also implicitly assume that the derivatives $\partial_i:=\partial_{\xi^i}=\frac{\partial}{\partial\xi^i}$ $\forall i=1,\ldots,d$ commute with integration.} Following \cite{amari2000methods}, we furthermore call a family of probability distributions 
\begin{align}
\mathcal{S}=\{p(X,\xi)|X\in\mathbb{V} \text{ and }\xi\in \Xi\}\,,\label{DefStatModel}
\end{align}
such that $\xi\to p(X,\xi)$ is injective $\forall X\in\mathbb{V}$, a \emph{statistical model}. It was shown in \cite{amari2000methods} that $\mathcal{S}$ can be endowed with the structure of a differentiable manifold (also called \emph{statistical manifold}) and equipped with a Riemannian metric, called the \emph{Fisher information metric}\footnote{When simply viewed as a $d\times d$ matrix, this quantity is also called the \emph{Fisher information matrix} or in the case $d=1$ simply the \emph{Fisher information}.} \cite{Rao,Jeffreys,BURBEA1982575,amari2000methods,Nielsen}
\begin{align}
g_{ij}(\xi):&=\sum_{X\in\mathbb{V}}\left(\partial_i\log p(X,\xi)\right)\left(\partial_j\log p(X,\xi)\right)\,p(X,\xi)\hspace{2.2cm}\forall i,j\in\{1,\ldots,d\}\,,\nonumber\\
&=-\sum_{X\in\mathbb{V}}\left(\partial_i\partial_j\,\log p(X,\xi)\right)\,p(X,\xi)=4\sum_{X\in\mathbb{V}}\left(\partial_i\sqrt{p(X,\xi)}\right)\left(\partial_j\sqrt{p(X,\xi)}\right)\,.\label{DefGenMetric}
\end{align}
For a given $\mathcal{S}$, this metric can be used to define a distance between two probability distributions~\cite{Rao,Jeffreys} (see also \cite{Nielsen,Miyamoto}): let $\xi_c:\,[0,1]\to \Xi$ be a curve in $\Xi$ (parametrised by $\tau\in[0,1]$). 
Such a curve is a geodesic if it satisfies
\begin{align}
&\ddot{\xi}_c^k+{\Gamma^k}_{ij}\,\dot{\xi}_c^i\,\dot{\xi}_c^j=0\,,&&\text{with} &&\begin{array}{l}\dot{\xi}^i_c=\frac{d\xi_c^i}{d\tau}\,,\\ {\Gamma^k}_{ij}=\frac{1}{2}\,g^{kl}\left(\partial_i g_{jl}+\partial_j g_{li}-\partial_l g_{ij}\right)\,,\end{array}
\end{align}
where $g^{ij}$ is the inverse of the metric (\ref{DefGenMetric}). We can then define the \emph{Fisher-Rao distance} of the probability distributions $p(X,\xi_1)$ and $p(X,\xi_2)$ for $\xi_{1,2}\in \Xi$ as 
\begin{align}
d(\xi_1,\xi_2)=\int_0^1 d\tau\,\sqrt{\dot{\xi}^i_c\,\dot{\xi}^j_c\,g_{ij}(\xi_c)}\,,\label{DefFisherRaoDistance}
\end{align}
where $\xi_c$ is the geodesic such that $\xi_c(\tau=0)=\xi_1$ and $\xi_c(\tau=1)=\xi_2$.


\subsection{Time Evolution and Flow Equations}\label{Sect:FixedPointDynamics}
In this paper, we shall exclusively be interested in the one-dimensional case $\Xi\subset\mathbb{R}$, where we shall interpret $t\in\Xi$ as a time variable. In this case, we use the following notation for the one-dimensional metric (following from the general definition (\ref{DefGenMetric})).
\begin{align}
&d\mathfrak{p}^2(t)=g_{tt}(t)\,dt^2\,,&&\text{with} &&g_{tt}(t)=\sum_{X\in\mathbb{V}}\left(\frac{\partial \log p(X,t)}{\partial t}\right)^2\,p(X,t)\,,\label{DefFisherInformationMatrix}
\end{align}
with $dt$ an infinitesimal time element.\footnote{As remarked before, $g_{tt}$ is the Fisher information, originally introduced in \cite{Fisher}. In the following, to highlight its property to keep track of changes or differences of $p(X,t)$, we shall refer to it as 'Fisher (information) metric' (or simply 'metric' for short).} We first remark that the metric element $d\mathfrak{p}^2(t)$ can in fact be related to the Kullback-Leibler divergence (\ref{DefKullbackLeibler}) of $p(X,t)$ and $p(X,t+dt)$ (see \cite{lesne_2014}))
\begin{align}
d\mathfrak{p}^2(t)=2\,D\left(p(\cdot,t)||p(\cdot,t+dt),t\right)\,.
\end{align}
Furthermore, the Fisher-Rao distance (\ref{DefFisherRaoDistance}) becomes the (absolute value) of the proper time
\begin{align}
d(t_1,t_2)=\left|\int_{t_1}^{t_2}\sqrt{g_{tt}(t')}\,dt'\right|\,.\label{FisherRaoOneDim}
\end{align}
Moreover, we shall argue that (\ref{DefFisherInformationMatrix}) can be used to describe the time evolution of the probability distribution $p(X,t)$, at least in certain cases. To this end, we focus on a finite set $\mathbb{V}=\{X_1,\ldots,X_{N+1}\}$ of cardinality $N+1$ and define the probability distribution
\begin{align}
p:\,\mathbb{V}\times \mathbb{R}&\longrightarrow [0,1]\,,&&\text{with} &&\sum_{i=1}^{N+1} p_i(t)=1\,,\nonumber\\
(X_i,t)&\longmapsto p(X_i,t)=p_i(t)\,. \label{DefProbabilityDistributions}
\end{align}
We resolve the normalisation condition by defining $(x_1,\ldots,x_N)$ (with $x_i(t)\in[0,1]$), such that
\begin{align}
&p_i(t)=x_i(t)\hspace{0.5cm}\forall i\in\{1,\ldots, N\}\,,&&\text{and} &&p_{N+1}(t)=1-\sum_{i=1}^N x_i(t)\,,
\end{align}
The Fisher information metric (\ref{DefFisherInformationMatrix}) can then be written as (with $\dot{x}_i=\frac{dx_i}{dt}$)
\begin{align}
g_{tt}=\sum_{i=1}^{N+1}p_i(t)\,\left(\frac{d}{dt}\,\log p_i(t)\right)^2=\sum_{i=1}^N\frac{\left(1-\sum_{j\neq i}^N x_j\right)\,\dot{x}_i^2}{x_i\left(1-\sum_{j=1}^N x_j\right)}+2\sum_{1\leq i<j\leq N}\frac{\dot{x}_i \dot{x}_j}{1-\sum_{k=1}^N x_k}\,.\label{MetEq}
\end{align}
Rather than calculating the metric for a (given) probability distribution, we shall in the following try to extract information about $\dot{x}_i$ in terms of the metric $g_{tt}$, taking the latter as an input. In general, it is not possible to extract the time evolution for all $x_i$ in this manner, however, by making further assumptions on the form of the dynamics, we can derive certain insights. As an example, consider the case in which one of the probabilities is very close to zero, \emph{e.g.} without loss of generality $x_1\ll 1$. In this case, the metric (\ref{MetEq}) is dominated by a single term
\begin{align}
&x_1\ll 1:&&g_{tt}=\frac{\dot{x}_1^2}{x_1}+\mathcal{O}(x_1^0)\,,&&\text{such that}&&\frac{dx_1}{dt}\sim\pm\sqrt{x_1\,g_{tt}}\,.\label{EffectiveSmallX}
\end{align}
The sign in the last equation cannot be uniquely fixed due to time inversion symmetry and is determined whether the probability $x_1$ is growing or diminishing. Notice, that (\ref{EffectiveSmallX}) is an effective description of the time evolution of $x_1$, where the entire dynamics of the rest of the system (\emph{i.e.} $x_{2,\ldots,N}$) is encoded in the metric $g_{tt}$.

In the remainder of this paper, we shall study in detail a different scenario than (\ref{EffectiveSmallX}): consider the case where the time derivative of one probability is much larger than the others\footnote{From the perspective of the $p_i$, the condition (\ref{DefOneDimFlow}) implies that $\dot{x}_1$ grows (diminishes) at the expense of $p_{N+1}$, while the remaining probabilities remain essentially constant. This scenario therefore corresponds to a competition between two probabilities, which can both take values in the interval $[0,\alpha]$.}, thus, without loss of generality
\begin{align}
&|\dot{x}_1|\gg |\dot{x}_i|\sim 0\hspace{0.5cm} \forall i\in\{2,\ldots,N\}\,,&&\text{such that} &&\sum_{i=2}^N x_i(t)\sim 1-\alpha=\text{const}.\,.\label{DefOneDimFlow}
\end{align}
In this case, the metric can be approximated by
\begin{align}
g_{tt}\sim\frac{\left(1-\sum_{j=2}^N x_j\right)}{x_1\left(1-\sum_{j=1}^N x_j\right)}\,\dot{x}_1^2+\mathcal{O}(\dot{x}_1)=\frac{\alpha\, \dot{x}_1^2}{x_1(\alpha-x_1)}+\mathcal{O}(\dot{x}_1)\,.\label{MetricFormFirst}
\end{align}
This equation can be re-written in the following fashion
\begin{align}
&\frac{dx_1}{dt}\sim\pm\sqrt{g_{tt}\,x_1\,\left(1-\frac{x_1}{\alpha}\right)}\,,&&\text{with} &&x_1\in[0,\alpha]\,,\label{FlowEquation}
\end{align} 
where the sign cannot be fixed and is determined by whether the probability is growing or diminishing. If $g_{tt}$ is considered as an (external) input, equation (\ref{FlowEquation}) can be understood as an effective description of the dynamics for $x_1$. In particular, in the case that $g_{tt}$ is given as a function of $x_1$ (thereby encoding the influence of the remainder of the system on $x_1$), (\ref{FlowEquation}) uniquely fixes the time evolution of $x_1$. In this case, (\ref{FlowEquation}) plays a similar role to the flow equations in epidemiology discussed in \cite{,Cacciapaglia:2021cjl}. Notice in this respect that $x_1=0$ and $x_1=\alpha$ denote the extremal values $x_1$ can take. 

We furthermore remark, that in the particular case (\ref{DefOneDimFlow}) the dynamics of $x_1$ can in fact be solved in terms of the proper time, which is the Fisher-Rao distance (\ref{DefFisherRaoDistance}) in its form (\ref{FisherRaoOneDim}). To this end, we rewrite the flow equation (\ref{FlowEquation})
\begin{align}
\frac{dx_1}{\sqrt{x_1\left(1-\frac{x_1}{\alpha}\right)}}=\pm\sqrt{g_{tt}}\,dt\,.
\end{align}
Integrating this equation, with initial condition $x_1(t_0)=x_0$ (with $t>t_0$), we obtain
\begin{align}
d(t,t_0)=\pm 2\sqrt{\alpha}\,\left[\arcsin\left(\sqrt{{x_1}/\alpha}\right)-\arcsin\left(\sqrt{x_0/\alpha}\right)\right]
\end{align}
which can be inverted to give
\begin{align}
x_1(t)=\alpha\sin^2\left[\arcsin\left(\sqrt{\tfrac{x_0}{\alpha}}\right)\pm \tfrac{d(t,t_0)}{2\sqrt{\alpha}}\right]\,,
\end{align}
where the sign determines whether $x_1$ is growing or decreasing as a function of $d$.


In the following Sections~\ref{Sect:SimpleModels} we shall consider simple models in epidemiology and population dynamics and show that their dynamics can indeed be re-written in the form (\ref{FlowEquation}). In particular, we shall provide an explicit form of the metric $g_{tt}$ in terms of $x_1$ (either analytically or, in more complicated cases, numerically). Based on these examples, we shall propose in Section~\ref{Sect:GeneralMetric} a unified description of the various examples discussed in Sections \ref{Sect:SimpleModels}, which in its simplest form allows a completely analytic solution.

\section{Information Theory in Simple Theoretical Models}\label{Sect:SimpleModels}
In this Section we first discuss as simple examples the Lotka-Volterra model~\cite{lotka_alfred_j_1909_1428768,Volterra} in Subsection~\ref{Sect:LoktaVolterra} and the SIR-model~\cite{Kermack:1927} in Subsection~\ref{Sect:EpidemiologyModels}. The latter shall be generalised in Subsections~\ref{Sect:SInModel} and \ref{Sect:SIIRModel} to include multiple variants of a pathogen.


\subsection{Lotka-Volterra Model}\label{Sect:LoktaVolterra}
\begin{figure}[htbp]
\begin{center}
\includegraphics[width=7.5cm]{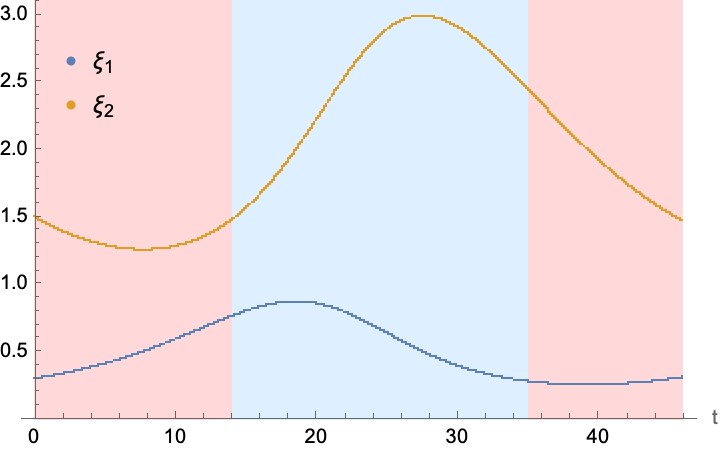}\hspace{1cm}\includegraphics[width=7.5cm]{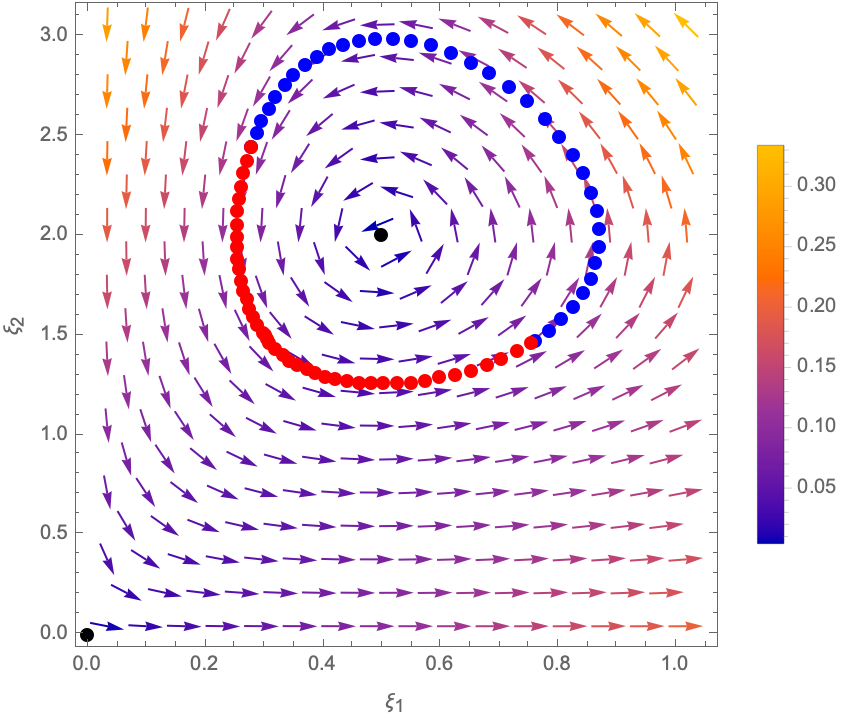}
\end{center}
\caption{Left panel: numerical solution of (\ref{DefLotkaVolterraEq}) for $a_1=0.2$, $a_2=0.1$, $b_1=0.1$, $b_2=0.2$ and initial conditions $\xi_1(0)=0.3$, $\xi_2(0)=1.5$. Right panel: trajectory of the system (red and blue points) in the $(\xi_1,\xi_2)$-plane. The arrows indicate the size and direction of the derivatives of $\xi_{1,2}$ (following from (\ref{DefLotkaVolterraEq})) and the black dots denote the fixed points of the system at $(\xi_1,\xi_2)=(0,0)$ and $(\xi_1,\xi_2)=\left(\tfrac{a_2}{b_2},\tfrac{a_1}{b_1}\right)$. As explained in the text, in both panels the evolution of the system is separated into regions with monotonically growing (red) or decreasing (blue)~$x$ (see also left panel of Figure~\ref{Fig:LotkaVolterraEntropyMetric}). }
\label{Fig:LotkaVolterraSol}
\end{figure}

A simple model for population dynamics are the Lotka-Volterra equations. They describe the time evolution of two species, usually referred to as 'prey' and 'predator', whose population growth depends on each other: denoting the respective size of their population (as a function of time $t\in\mathbb{R}_+$) by $\xi_{1,2}:\,\mathbb{R}_+\to\mathbb{R}_+$, the dynamics is described by
\begin{align}
&\frac{d\xi_1}{dt}=a_1\,\xi_1-b_1\,\xi_1\,\xi_2\,,&&\frac{d\xi_2}{dt}=-a_2\,\xi_2+b_2\,\xi_1\,\xi_2\,,&&\text{with} &&a_{1,2}\,, b_{1,2}\in\mathbb{R}_+\,.\label{DefLotkaVolterraEq}
\end{align}
In short, the population of prey animals ($\xi_1$) grows exponentially on its own, but is limited by the presence of the predators ($\xi_2$). The latter in turn decline exponentially on their own and only grow depending on the available prey. An exemplary numerical solution for $(\xi_1,\xi_2)$ is shown in Figure~\ref{Fig:LotkaVolterraSol}: the solutions never reach one of the equilibrium points at $(\xi_1,\xi_2)=(0,0)$ or $(\xi_1,\xi_2)=\left(\tfrac{a_2}{b_2},\tfrac{a_1}{b_1}\right)$ but are in fact periodic, as can be seen from the trajectory of the system in the $(\xi_1,\xi_2)$-plane in the right panel of Figure~\ref{Fig:LotkaVolterraSol}.

\begin{figure}[htbp]
\begin{center}
\includegraphics[width=7.5cm]{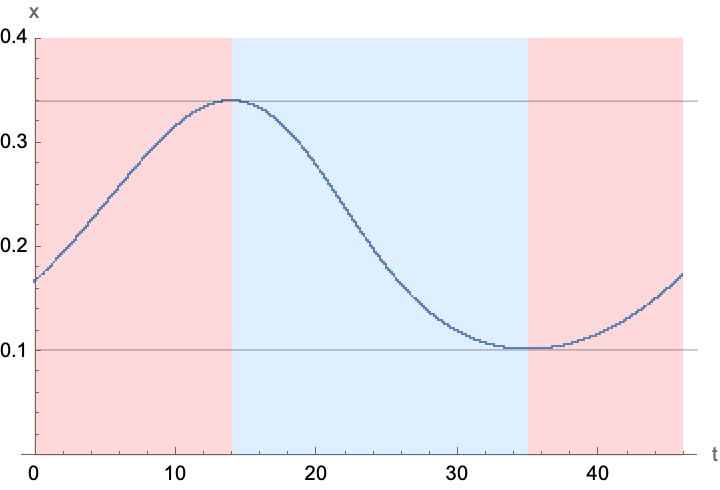}\hspace{1cm}\includegraphics[width=7.5cm]{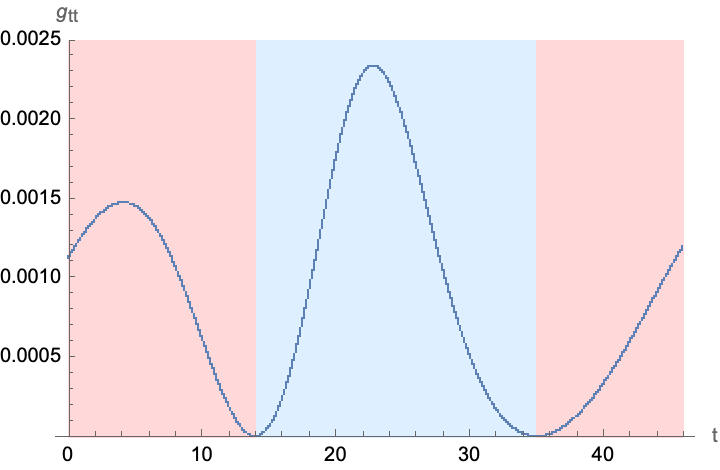}
\end{center}
\caption{Left panel: Numerical plot of the probability $x$ as a function of time, which is bounded in the intervall $x\in[0.102,0.304]$. Right panel: the Fisher information metric (\ref{LVFisherMetric}). Both plots use the same numerical values as in Figure~\ref{Fig:LotkaVolterraSol} and follow the same colouring scheme: the extrema of $x$ correspond to zeroes of the metric.}
\label{Fig:LotkaVolterraEntropyMetric}
\end{figure}

To facilitate the further discussion, we have divided the trajectory in Figure~\ref{Fig:LotkaVolterraSol} into two parts (coloured red and blue respectively): to explain this, we first define the probability distribution $p:\,\{1,2\}\times \mathbb{R}\to [0,1]$ with
\begin{align}
&p_1=\frac{\xi_1}{\xi_1+\xi_2}=:x\,,&&p_2=\frac{\xi_2}{\xi_1+\xi_2}=1-x\,.\label{DefProbs}
\end{align}
The probability $p_1(t)=x(t)$ (which depends on time $t$) takes values only in a finite interval and has extrema for
\begin{align}
a_1+a_2=b_1 \xi_2+b_2\xi_1\,.
\end{align}
On the blue portion of the trajectory, $x$ is monotonically decreasing, while on the red portion of the trajectory, $x$ is monotonically increasing, as can be seen in the left panel of Figure~\ref{Fig:LotkaVolterraEntropyMetric}. The respective regimes in the time-evolution of the system are also indicated in red and blue respectively in the left panel of Figure~\ref{Fig:LotkaVolterraSol} (and all subsequent Figures).

\begin{figure}[htbp]
\begin{center}
\includegraphics[width=7.5cm]{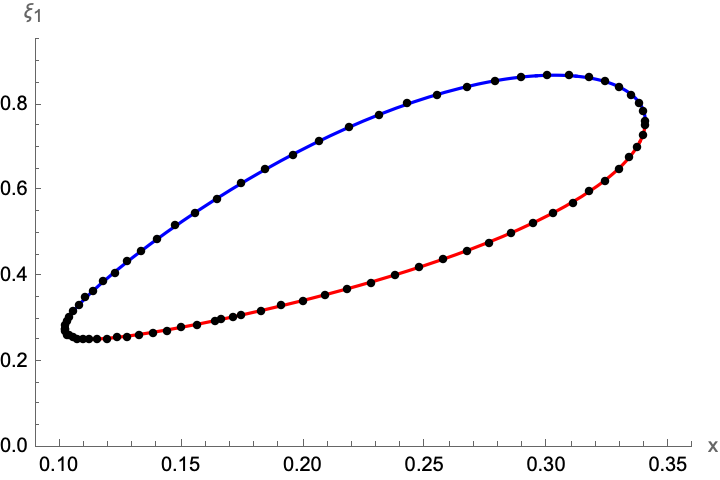}\hspace{1cm}\includegraphics[width=7.5cm]{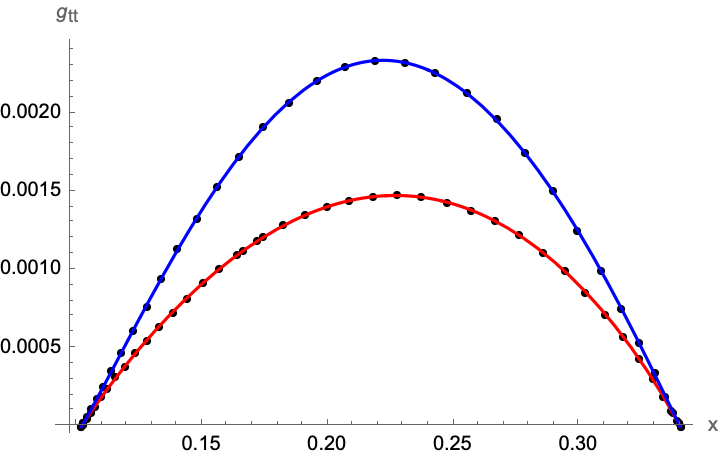}
\end{center}
\caption{Left panel: solution $\xi_1$ as a function of $x$ in (\ref{Formx1}). Right panel: metric (\ref{AnalyticMetricLV}) as a function of $x$. In both cases, the red and blue curve represent the two branches of the Lambert function, while the black dots represent a numerical solution. The parameters were chosen to be $a_1=0.2$, $a_2=0.1$, $b_1=0.1$, $b_2=0.2$ and initial conditions $\xi_1(0)=0.3$, $\xi_2(0)=1.5$ (such that $L\sim 0.249$).}
\label{Fig:LotkaVolterraAnalytic}
\end{figure}

\begin{figure}[htbp]
\begin{center}
\includegraphics[width=7.5cm]{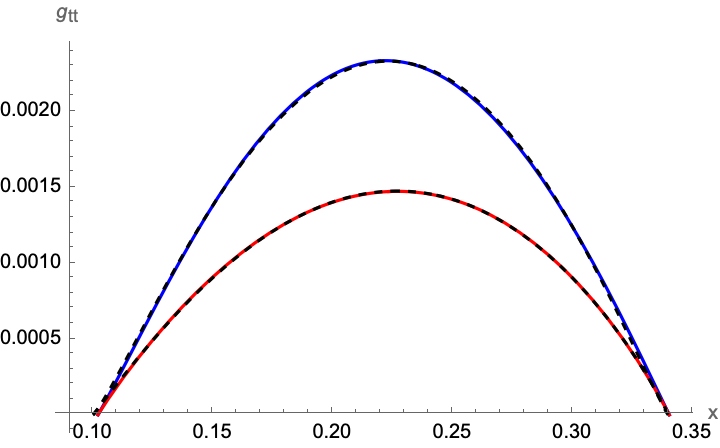}\hspace{1cm}\includegraphics[width=7.5cm]{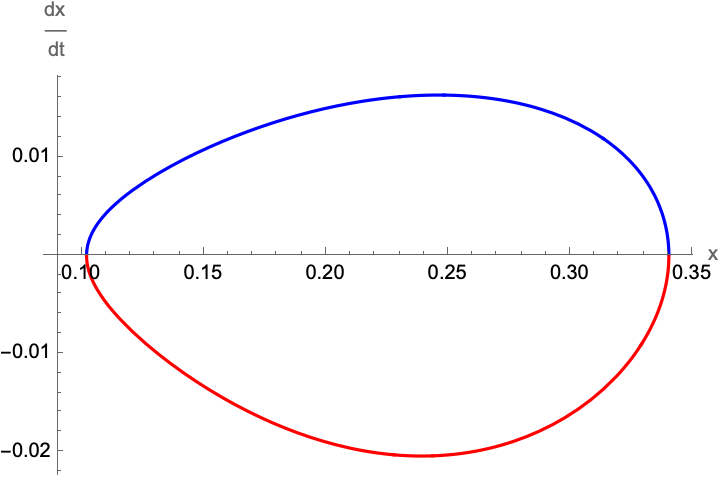}
\end{center}
\caption{Left panel: Approximation of the metric $g_{tt}$ in (\ref{AnalyticMetricLV}) as a function of $x$, using the same numerical values as in Figure~\ref{Fig:LotkaVolterraAnalytic}. The black dashed lines are interpolations using the functions (\ref{LVinter}), with $x_1\sim 0.1$, $x_2\sim 0.34$ and $(a,b,c)\sim (0.31,1.19,1.13)$ for the blue curve and $(a,b,c)\sim (0.09,1.05,0.93)$ for the red curve. Right panel: time derivative of $x$, following the flow equation (\ref{FlowEquation}) for $x$ computed from the metric $g_{tt}$.}
\label{Fig:LotkaVolterraFlow}
\end{figure}

Using the probabilities (\ref{DefProbs}), we can calculate the Fisher information metric (\ref{DefFisherInformationMatrix}) (which is plotted as a function of time in the right panel of Figure~\ref{Fig:LotkaVolterraEntropyMetric}): 
\begin{align}
g_{tt}=(\partial_t \log(p_1))^2\, p_1+(\partial_t \log(p_2))^2\, p_2=\frac{\xi_1 \xi_2(a_1+a_2-b_2\xi_1-b_1\xi_2)^2}{(\xi_1+\xi_2)^2}\,.\label{LVFisherMetric}
\end{align}
Following (\ref{FlowEquation}), extrema of the probability $x$ correspond to zeroes of the Fisher metric. Instead of writing the latter as a function of time $t$, we shall now re-write it as a function of $x$. To this end, as a first step, we can express $\xi_2$ as a function of $x$ and $\xi_1$ to obtain
\begin{align}
g_{tt}=\frac{1-x}{x}\,\left(a_1x+a_2x-\left(b_1(1-x)+b_2x\right)\xi_1\right)^2\,.\label{LVFisherPre}
\end{align}
However, to further express $\xi_1$ in terms of $x$, we need to solve the dynamics of the system by integrating the quotient of the two equations in (\ref{DefLotkaVolterraEq})
\begin{align}
\left(b_2+b_1\frac{1-x}{x}\right) \xi_1-(a_1+a_2)\,\log\xi_1-a_1\log\left(\frac{1-x}{x}\right)=L\,,\label{InitCond}
\end{align}
where $L=b_2\xi_1(0)-a_2\log\xi_1(0)+b_1\xi_2(0)-a_1\log\xi_2(0)$ is a constant that is determined by the initial conditions. Equation (\ref{InitCond}) can be solved to express $\xi_1$ as a function of $x$
\begin{align}
\xi_1=-\frac{(a_1+a_2)x}{b_1(1-x)+b_2 x}\,W\left(-\frac{e^{-\frac{L}{a_1+a_2}}\left(\frac{1-x}{x}\right)^{-\frac{a_1}{a_1+a_2}}(b_1(1-x)+b_2 x)}{x(a_1+a_2)}\right)\,,\label{Formx1}
\end{align}
where $W$ is the Lambert function. Notice that (\ref{Formx1}) are in fact two solutions, corresponding to the two branches of the Lambert function, as is schematically shown in the left panel of Figure~\ref{Fig:LotkaVolterraAnalytic}. These two branches are again related to the sign of $\frac{dx}{dt}$ (as indicated by the colouring of the plot in Figure~\ref{Fig:LotkaVolterraAnalytic}). The Fisher information metric (\ref{LVFisherPre}) then takes the form
\begin{align}
g_{tt}&=(a_1+a_2)^2 \,x\,(1-x)\left[1+W\left(-\frac{e^{-\frac{L}{a_1+a_2}}\left(\frac{1-x}{x}\right)^{-\frac{a_1}{a_1+a_2}}(b_1(1-x)+b_2 x)}{x(a_1+a_2)}\right)\right]^2\,,\label{AnalyticMetricLV}
\end{align}
which (as (\ref{Formx1})) has two branches, as can be seen from the right panel of Figure~\ref{Fig:LotkaVolterraAnalytic}. The zeroes of the metric are the solutions of the equation
\begin{align}
b_1\left(\frac{1-x}{x}\right)^{\frac{a_2}{a_1+a_2}}+b_2\,\left(\frac{1-x}{x}\right)^{-\frac{a_1}{a_1+a_2}}=(a_1+a_2)\,e^{\frac{L}{a_1+a_2}-1}\,,\label{EqSolutionsZeroes}
\end{align}
which cannot be solved analytically for generic values of $a_{1,2}$. We note, however, that the form of the metric (\ref{AnalyticMetricLV}) can be very well approximated by a function of the form
\begin{align}
&g_{tt}(x)=a\,(x-x_1)^b\,(x_2-x)^c\,,&&\forall x\in[x_1,x_2]\,.\label{LVinter}
\end{align}
where $x_{1,2}\in[0,1]$ (with $x_1<x_2$) are two real solutions of (\ref{EqSolutionsZeroes}) (which are independent of time $t$) and $a,b,c$ suitable fitting coefficients of order 1. Indeed, an approximation of this form is shown in Figure~\ref{Fig:LotkaVolterraFlow}, along with the flow equation 
\begin{align}
&\frac{dx}{dt}=\pm\sqrt{g_{tt}\,x\,\left(1-x\right)}\,,
\end{align} 
following from (\ref{FlowEquation}) (with $\alpha=1$).

\begin{figure}[htbp]
\begin{center}
\includegraphics[width=7.5cm]{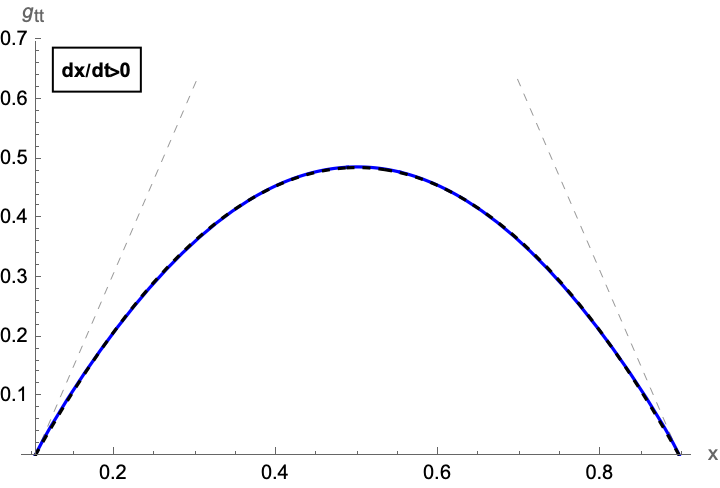}\hspace{1cm}\includegraphics[width=7.5cm]{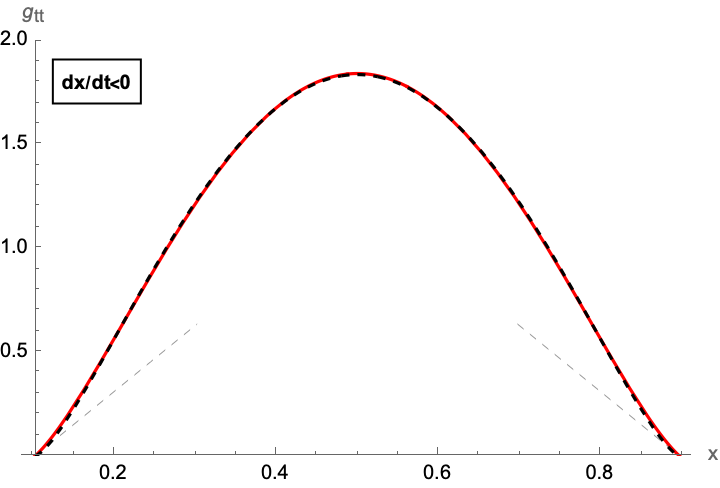}\\
\end{center}
\caption{The two branches of the metric (\ref{AnalyticMetricLVachoice}) for $L=3$ (left: blue solid line, right: red solid line). The dashed black lines indicate the best interpolation with $(\kappa,d)=(0.765,0.997)$ (left panel) and $(\kappa,d)=(6.119,1.403)$. The dashed straight lines show the tangents at the two zeroes of the metric.}
\label{Fig:LotkaVolterraAnalyticMetric}
\end{figure}

To make this approximation more tangible, we shall consider a concrete example, namely\footnote{By defining $\xi_1=\frac{a_2}{b_2}\,\Xi_1$, $\xi_2=\frac{a_2}{b_1}\,\Xi_2$ and $t=\tau/a_2$, the equations (\ref{DefLotkaVolterraEq}) can always be brought into the form
\begin{align}
&\frac{d\Xi_1}{d\tau}=\alpha\,\Xi_1-\Xi_1\,\Xi_2\,,&&\text{and} &&\frac{d\Xi_2}{d\tau}=-\Xi_1+\Xi_1\,\Xi_2\,.
\end{align}
with $\alpha=a_1/a_2$. (\ref{RedLotkaVoltera}) is therefore the particular choice $\alpha=1$.}
\begin{align}
a_1=a_2=b_1=b_2=1\,.\label{RedLotkaVoltera}
\end{align}
In this case, relation (\ref{InitCond}) allows for real solutions $\xi_1$ in terms of $x$ only for $L\geq 2$. Furthermore, (\ref{EqSolutionsZeroes}) has two real solutions in the interval $[0,1]$\footnote{For $L=2$ we have $x_1=x_2=1/2$. Furthermore, the only real solutions of (\ref{InitCond}) arise for $x=0$ or $x=1/2$, in which case, the system is in one of the two equilibrium points. In the following, we shall therefore implicitly assume $L>2$.}
\begin{align}
&x_1=\frac{1}{2}\,\left(1-\sqrt{1-e^{2-L}}\right)\,,&&x_2=\frac{1}{2}\,\left(1+\sqrt{1-e^{2-L}}\right)\,,&&\forall L> 2\,.
\end{align}
Furthermore, the metric (\ref{AnalyticMetricLV}) takes the simpler form
\begin{figure}[htbp]
\begin{center}
\includegraphics[width=7.5cm]{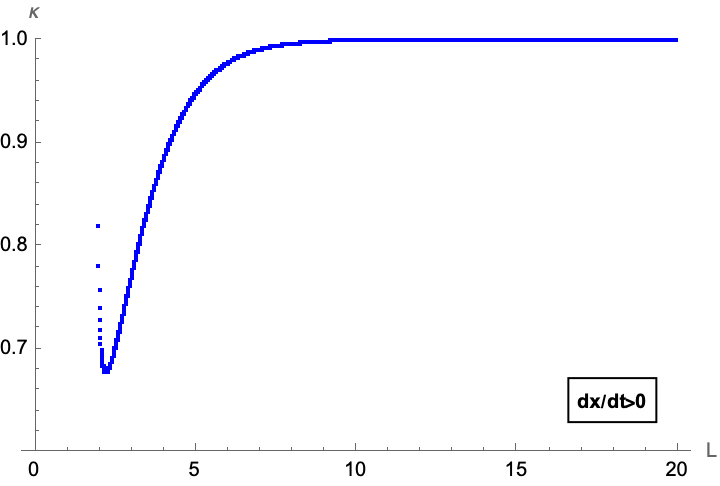}\hspace{1cm}\includegraphics[width=7.5cm]{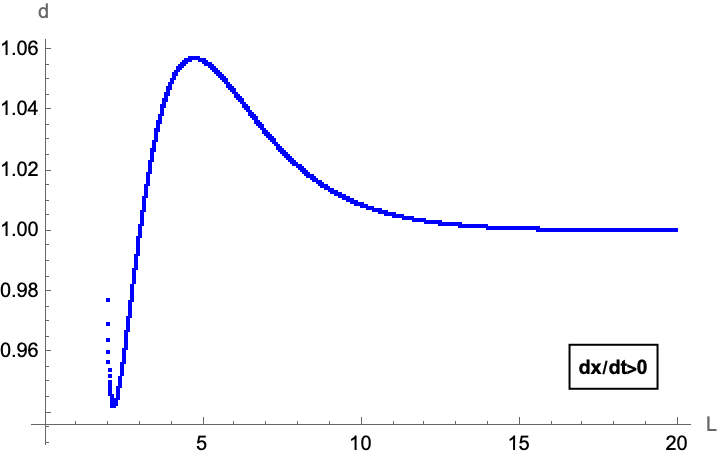}\\
\includegraphics[width=7.5cm]{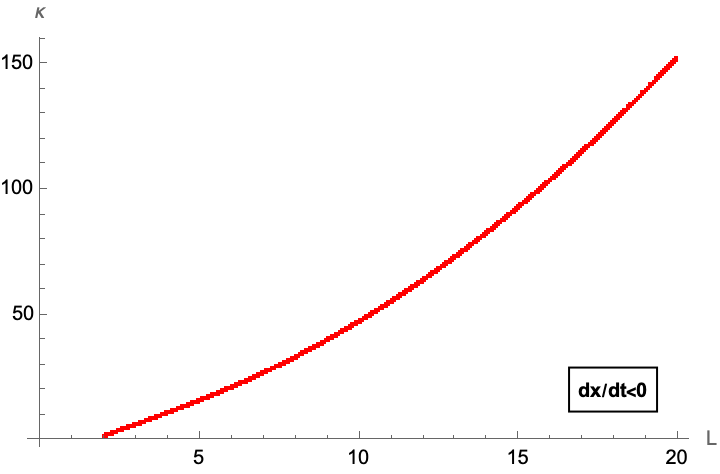}\hspace{1cm}\includegraphics[width=7.5cm]{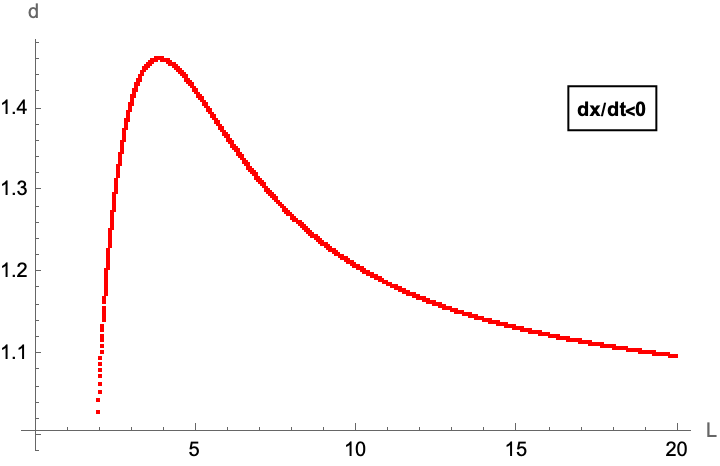}
\end{center}
\caption{Best fit of the parameters $\kappa$ and $d$ in (\ref{AnalyticMetricLVapprox}) as functions of $c$. The top panels show the case $\frac{dx}{dt}>0$ and the top panels $\frac{dx}{dt}>0$, corresponding to two different branches of the Lambert function in (\ref{AnalyticMetricLVachoice}).}
\label{Fig:LotkaVolterraAnalyticFitc}
\end{figure}
\begin{align}
g_{tt}=4 \,x\,(1-x)\left[1+W\left(-\frac{e^{-L/2}}{2\sqrt{x(1-x)}}\right)\right]^2\,,\label{AnalyticMetricLVachoice}
\end{align}
which can be approximated by
\begin{align}
g_{tt}\sim 4\kappa\left(x-\frac{1}{2}\,\left(1-\sqrt{1-e^{2-L}}\right)\right)^d\,\left(\frac{1}{2}\,\left(1+\sqrt{1-e^{2-L}}\right)-x\right)^d\,,\label{AnalyticMetricLVapprox}
\end{align}
where $(\kappa,d)$ are fit parameters that depend on $L$: The best fits are plotted schematically in Figure~\ref{Fig:LotkaVolterraAnalyticFitc}, while explicit plots of the metric (for $L=3$) are shown in Figure~\ref{Fig:LotkaVolterraAnalyticMetric}.
 
\subsection{SIR(S) Model}\label{Sect:EpidemiologyModels}
We next consider a simple compartmental model in epidemiology that describes the spread of a pathogen through a population. Here, we only consider the spread of a single pathogen, while in the following Subsections, we shall consider multiple different variants at the same time.

\begin{figure}[htbp]
\begin{center}
\includegraphics[width=7.5cm]{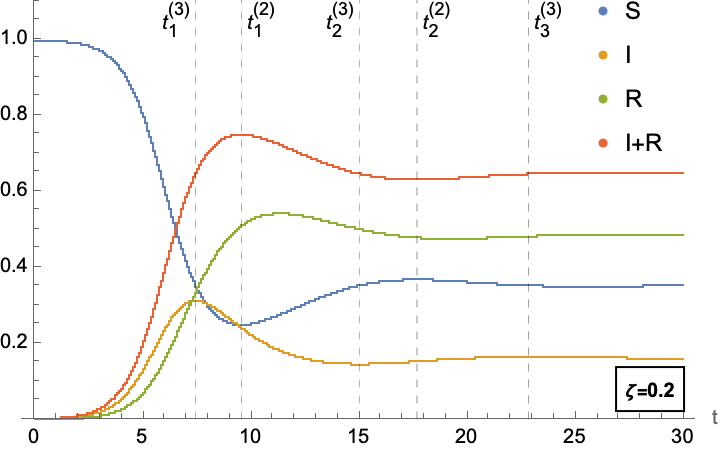}\hspace{1cm}\includegraphics[width=7.5cm]{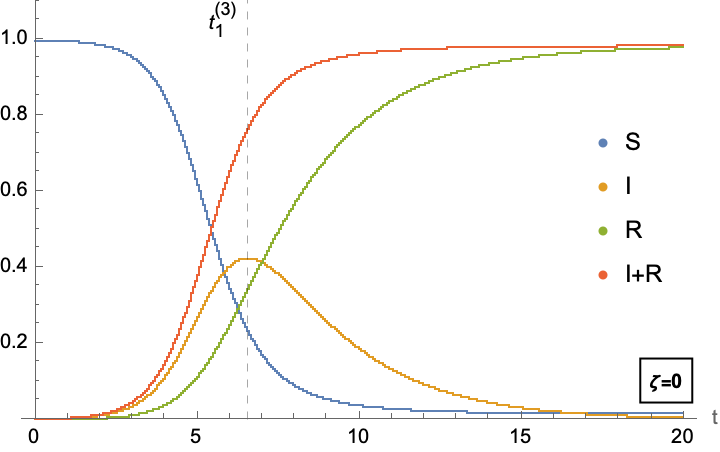}
\end{center}
\caption{Numerical solution for the SIR-model (\ref{SIRequations}) with and without reinfection: Left-panel: $\gamma=1.7$, $\sigma=0.6$, $\zeta=0.2$ and initial conditions $I(t=0)=0.00075$ (and $R(t=0)=0$). The times $t_{1,2}^{(2)}$ indicate extrema of $I+R$, while $t_{1,2,3}^{(3)}$ indicate extrema of $I$. Right panel: $\gamma=1.7$, $\sigma=0.4$, $\zeta=0$ and initial conditions $I(t=0)=0.00075$ (and $R(t=0)=0$). The time $t_1^{(3)}$ denotes the maximum of $I$.}
\label{Fig:SIRsol}
\end{figure}

We assume a(n isolated) population of constant size (normalised to 1), which we group into different compartments, regarding to their state of affliction with the pathogen: we denote with $S(t)$ the number of susceptible individuals, with $I(t)$ the number of infectious individuals and with $R(t)$ the number of removed individuals at time $t\in\mathbb{R}$.\footnote{Here we define as a susceptible an individual, who is currently not infectious (\emph{i.e.} not inflicted by the pathogen), but may become so if in contact with an infectious individual. In contrast, a removed individual is also not currently infectious but may also not become infectious upon contact with an infectious one.} We assume these numbers to be (differentiable) functions of time $S,I,R:\,\mathbb{R}\to [0,1]$, whose evolution is governed by~\cite{Kermack:1927} 
\begin{align}
&\frac{dS}{dt}=-\gamma\,S\,I+\zeta\,R\,,&&\frac{dI}{dt}=\gamma\,S\,I-\sigma I\,,&&\frac{dR}{dt}=\sigma\,I-\zeta R\,,\label{SIRequations}
\end{align}
and we choose initial conditions such that $(S+I+R)(t=0)=1$. Furthermore, $\gamma$, $\sigma$, $\zeta\in\mathbb{R}_+$ are constant parameters, which denote the rate at which susceptible individuals are infected, the rate at which infectious individuals become removed and the rate at which removed individuals become susceptible again\footnote{The model with $\zeta=0$ is usually referred to as the SIR model \cite{Kermack:1927}, while the model with $\zeta\neq 0$, \emph{i.e.} with the possibility of re-infection, is sometimes called the SIRS model (see \emph{e.g.} \cite{HethcoteThousand} for a review).}, respectively. A numerical solution of (\ref{SIRequations}) (distinguishing $\zeta\neq 0$ and $\zeta$=0) is shown in Figure~\ref{Fig:SIRsol}.

\begin{figure}[htbp]
\begin{center}
\includegraphics[width=7.5cm]{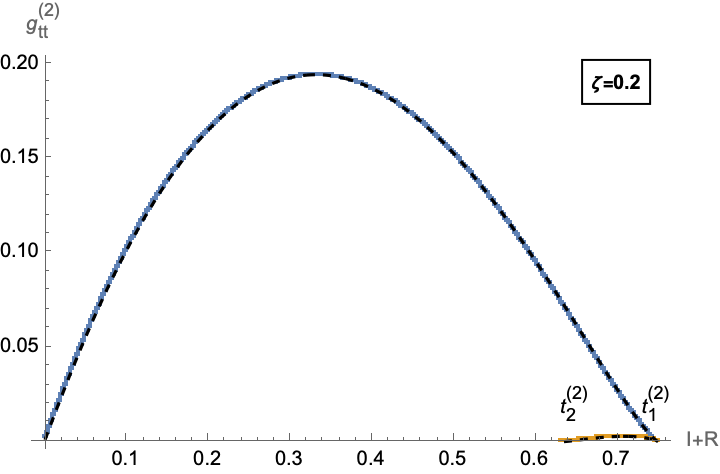}\hspace{1cm}\includegraphics[width=7.5cm]{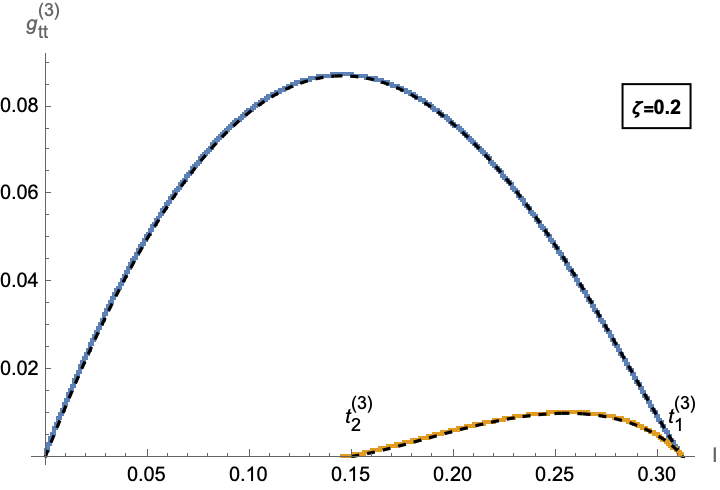}
\end{center}
\caption{Fisher metrics $g_{tt}^{(2)}$ as function of $I+R$ (left) and $g_{tt}^{(3)}$ as function of $I$ (right) in a model with reinfection. Both plots use the same parameters as in the left panel of Figure~\ref{Fig:SIRsol}. In the left panel, the blue curve shows the metric $g_{tt}^{(2)}$ from $t=0$ to $t=t_{1}^{(2)}$, where $d(I+R)/dt$ is positive (and thus the minus sign is applicable in (\ref{DynamicsMetric2})), while the orange curve shows the metric between $t=t_1^{(2)}$ and $t=t_{2}^{(2)}$, where $d(I+R)/dt$ is negative (and thus the plus sign is applicable in (\ref{DynamicsMetric2})). The dashed black lines represent interpolations of the form (\ref{LVinter}) with $(x_1,x_2,a,b,c)\sim (0,0.75,1.81,1.01,1.28)$ and $(x_1,x_2,a,b,c)\sim (0.63,0.75,1.11,1.41,0.74)$ respectively. In the right panel the blue curve shows the metric $g_{tt}^{(3)}$ from $t=0$ to $t=t_{1}^{(3)}$, where $dI/dt$ is positive (and thus the minus sign is applicable in (\ref{DynamicsMetric3})), while the orange curve shows the metric between $t=t_1^{(3)}$ and $t=t_{2}^{(3)}$, where $dI/dt$ is negative (and thus the plus sign is applicable in (\ref{DynamicsMetric3})). The dashed black lines represent interpolations of the form (\ref{LVinter}) with $(x_1,x_2,a,b,c)\sim (0,0.31,4.88,1.01,1.16)$ and $(x_1,x_2,a,b,c)\sim (0.14,0.31,2.99,1.56,0.79)$.}
\label{Fig:SIRshannonUnivrec}
\end{figure}

\begin{figure}[htbp]
\begin{center}
\includegraphics[width=7.5cm]{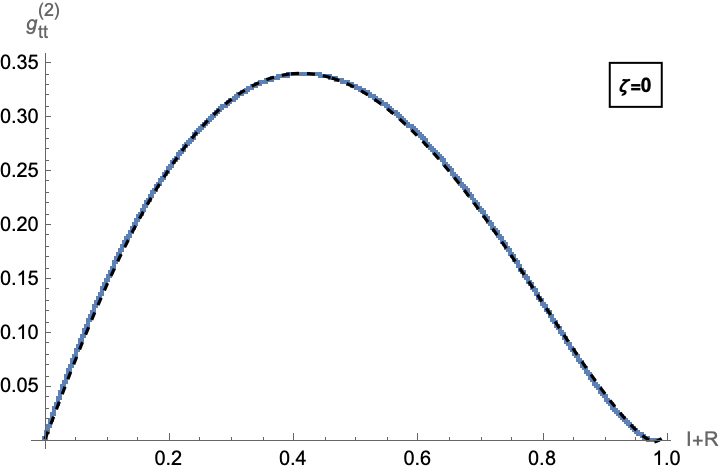}\hspace{1cm}\includegraphics[width=7.5cm]{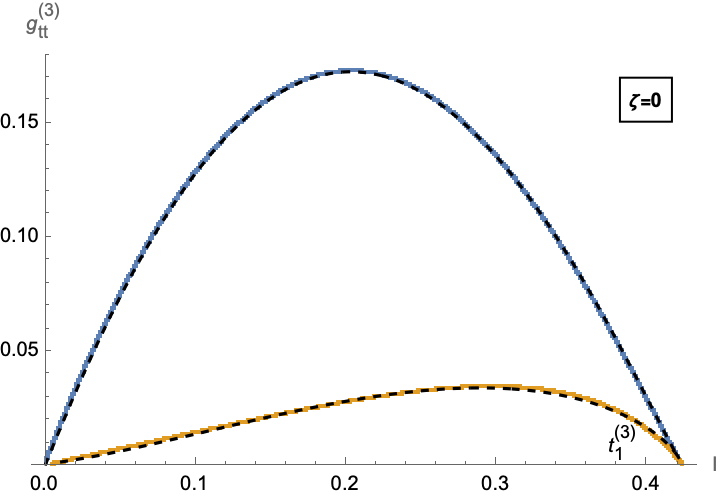}
\end{center}
\caption{Fisher metrics $g_{tt}^{(2)}$ as function of $I+R$ (left) and $g_{tt}^{(3)}$ as function of $I$ (right) in a model without reinfection. Both plots use the same parameters as in the right panel of Figure~\ref{Fig:SIRsol}. In the left panel, the dashed black line represents an interpolation of the form (\ref{LVinter}) with $(x_1,x_2,a,b,c)\sim (0,0.98,1.99,1.06,1.45)$. In the right panel the blue curve shows the metric $g_{tt}^{(3)}$ from $t=0$ to $t=t_{1}^{(3)}$, where $dI/dt$ is positive (and thus the minus sign is applicable in (\ref{DynamicsMetric3})), while the orange curve shows the metric between $t=t_1^{(3)}$ and $t=t_{2}^{(3)}$, where $dI/dt$ is negative (and thus the plus sign is applicable in (\ref{DynamicsMetric3})). The dashed black lines represent interpolations of the form (\ref{LVinter}) with $(x_1,x_2,a,b,c)\sim (0,0.42,4.68,1.02,1.11)$ and $(x_1,x_2,a,b,c)\sim (0,0.42,0.85,1.56,0.68)$ respectively.}
\label{Fig:SIRshannonUnivnonrec}
\end{figure}

Given the initial condition (and the fact that (\ref{SIRequations}) preserves the size of the population), such that $(S+I+R)(t)=1$ $\forall t\geq 0$, we can define different probability distributions that satisfy the normalisation condition (\ref{ProbNorm}). While a natural choice is to define (using the definition (\ref{DefProbabilityDistributions}))
\begin{align}
&p^{(1)}:\,\{1,2,3\}\times \mathbb{R}\to [0,1]&&\text{with}&&p_1^{(1)}(t)=S(t)\,,&&p_2^{(1)}(t)=I(t)\,,&&p_3^{(1)}(t)=R(t)\,,
\end{align}
in order to make contact with the formalism described in Section~\ref{Sect:FixedPointDynamics}, we shall be more interested in probability distributions as maps $\{1,2\}\times \mathbb{R}\to [0,1]$. Concretely, we shall consider two particular choices 
\begin{align}
&p^{(2)}:\,\{1,2\}\times \mathbb{R}\longrightarrow [0,1]&&\text{with}&&p_1^{(2)}(t)=S(t)\,,&&p_2^{(2)}(t)=(I+R)(t)\,,\nonumber\\
&p^{(3)}:\,\{1,2\}\times \mathbb{R}\longrightarrow [0,1]&&\text{with}&&p_1^{(3)}(t)=(S+R)(t)\,,&&p_2^{(3)}(t)=I(t)\,.\label{DefProbDefsSIR}
\end{align}
Notice, in the case $\zeta=0$, $p_2^{(2)}$ describes the cumulative number of infected individuals, which has been the object of study in the eRG approach (see \emph{e.g.} \cite{DellaMorte:2020wlc,DellaMorte:2020qry,cacciapaglia2020second,cacciapaglia2020mining,cacciapaglia2020evidence,cacciapaglia2020multiwave,Cacciapaglia:2020mjf,cacciapaglia2020us,Cacciapaglia:2021cjl,
Cacciapaglia:2021vvu,GreenPass,MLvariants}). For each of the probability distributions, we also introduce the Fisher information metrics (\ref{DefFisherInformationMatrix})
\begin{figure}[htbp]
\begin{center}
\includegraphics[width=7.5cm]{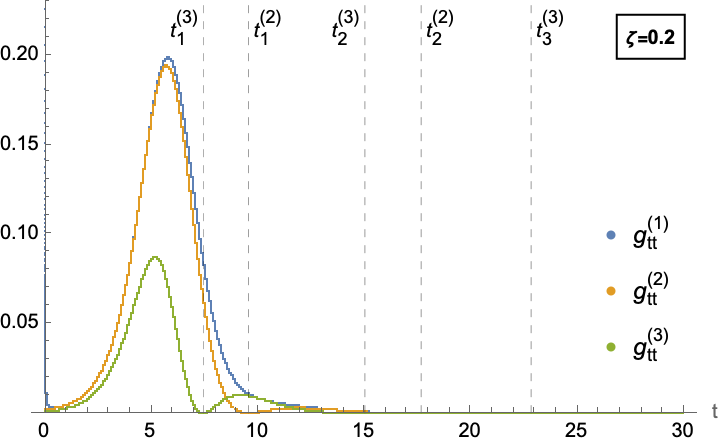}\hspace{1cm}\includegraphics[width=7.5cm]{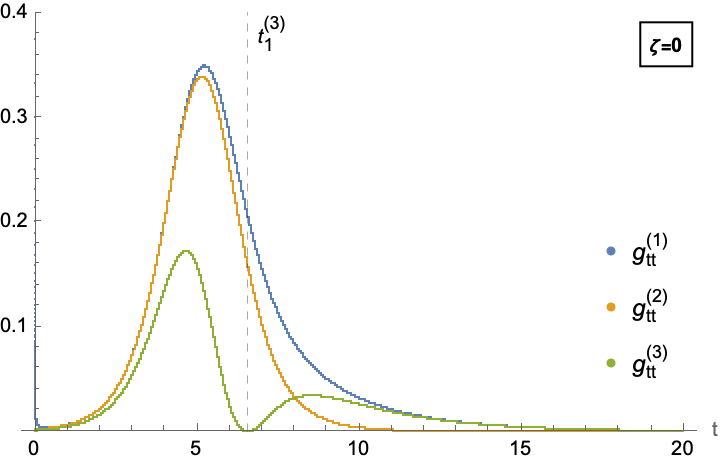}
\end{center}
\caption{Fisher metric associated with the probability distributions (\ref{DefProbDefsSIR}) for a SIR model with (left) and without (right) reinfection. The plots use the same numerical values as in Figure~\ref{Fig:SIRsol} and the vertical lines indicate extrema of $I+R$ and $I$ respectively.}
\label{Fig:SIRshannonRec}
\end{figure}
\begin{align}
g_{tt}^{(1)}(t)&=\sum_{i=1}^3 p_i^{(1)}\,(\partial_t \log p_i^{(1)})^2=\frac{(\gamma SI-\zeta R)^2}{S}+I(\gamma S-\sigma)^2+\frac{(\sigma I-\zeta R)^2}{R}\,,\nonumber\\
g_{tt}^{(2)}(t)&=\sum_{i=1}^2 p_i^{(2)}\,(\partial_t \log p_i^{(2)})^2=\frac{(\gamma SI-\zeta R)^2}{(I+R)S}\,,\nonumber\\
g_{tt}^{(3)}(t)&=\sum_{i=1}^2 p_i^{(3)}\,(\partial_t \log p_i^{(3)})^2=\frac{I(\gamma S-\sigma)^2}{R+S}\,,\label{FormMetricsGenSIR}
\end{align}
which are plotted as functions of time in Figure~\ref{Fig:SIRshannonRec}.\footnote{Notice that $g_{tt}^{(1)}$ (blue curve in Figure~\ref{Fig:SIRshannonRec}) is always large than $g_{tt}^{(2)}$ or $g_{tt}^{(3)}$ (orange and green curve). This follows on general grounds, since $p^{(2,3)}$ can be obtained as sums of $p^{(1)}$: indeed, following \cite{amari2000methods,AmariRao,AmariLoss} let $\mathbb{W}$, $\mathbb{V}$ be discrete sets and $F:\,\mathbb{V}\rightarrow \mathbb{W}$ a map, which allows to associate with each probability distribution $p(X,\xi)$ on $\mathbb{V}$ a probability distribution $q(F(X),\xi)$ on $\mathbb{W}$. If we denote by $g_{ij}^{(F)}$ the induced Fisher metric associated with $q(F(X),\xi)$, then $g_{ij}(\xi)-g_{ij}^{(F)}(\xi)$ is positive semidefinite.} Since $p^{(2)}$ and $p^{(3)}$ are probability distributions defined on a set $\mathbb{V}$ of cardinality $2$, the flow equation (\ref{FlowEquation}) (with $\alpha=1$) can be applied without any additional approximations (since (\ref{DefOneDimFlow}) is automatically satisfied).  Indeed, we can write
\begin{align}
&\frac{dS}{dt}=-\frac{d(I+R)}{dt}=\pm \sqrt{g_{tt}^{(2)}\,(R+I)S}\,,\label{DynamicsMetric2}\\
&\frac{d(S+R)}{dt}=-\frac{dI}{dt}=\pm \sqrt{g_{tt}^{(3)}\,(R+S)I}\,,\label{DynamicsMetric3}
\end{align}
where the sign can change during the different phases of the spread of the disease. In order for these equations to fully encode (part of) the dynamics of the system, the metrics $g_{tt}^{(2,3)}$ need to be expressed in terms of the probabilities themselves. This is shown for a numerical solution in Figure~\ref{Fig:SIRshannonUnivrec} for a model with reinfection and Figure~\ref{Fig:SIRshannonUnivnonrec} for a model without reinfection. We remark, although the model is not periodic, the schematic form of the metrics is similar to the shape found in Figure~\ref{Fig:LotkaVolterraFlow} in the Lotka Volterra model: indeed the dashed black lines give interpolations of all cuves with functions of the form (\ref{LVinter}). At least for the blue curves (which describe the initial phase of the epidemic, in which $I$ or $I+R$ are monotonically growing), the powers $b$ and $c$ are close to $1$.

For both $\zeta=0$ and $\zeta\neq 0$, the exact form of the metrics in Figures~\ref{Fig:SIRshannonUnivrec} and \ref{Fig:SIRshannonUnivnonrec} depends not only on the parameters $\gamma$, $\sigma$ and $\zeta$, but also on the exact initial conditions. In the case of $\zeta=0$ (\emph{i.e.} a model without re-infection) we can make this concrete by solving the dynamics analytically. To this end, we consider the two probability distributions $p^{(2)}$ and $p^{(3)}$ separately:
\begin{itemize}
\item metric $g_{tt}^{(2)}$: We start with the discussion of  the metric $g_{tt}^{(2)}$ in the second line of (\ref{FormMetricsGenSIR}). Let $x(t)=p_2^{(2)}(t)=(I+R)(t)$, such that for $\zeta=0$
\begin{align}
g_{tt}^{(2)}=\frac{\gamma^2}{x}\,(1-x)\,I^2\,,\label{FormPreExpgtt2}
\end{align}
which is not only a function of $x$ due to the presence of the factor $I^2$: in fact, precisely by replacing this factor by a function of $x$ we encode the non-trivial dynamics and turn (\ref{FlowEquation}) into a non-trivial flow equation (rather than a mere definition of $g_{tt}$). To obtain $I(x)$, we divide the second and third equations in (\ref{SIRequations})
\begin{align}
&\frac{dR}{dI}=\frac{d(x-I)}{dI}=\frac{\sigma}{\gamma(1-x)-\sigma}\,,&&\text{such that} &&\frac{dx}{dI}=\frac{\sigma}{\gamma(1-x)-\sigma}+1\,.\label{DiffSIRxdep2}
\end{align} 
This differential equation can be integrated by taking into account the initial conditions $x_0=x(t=0)=I(t=0)$ (since we are assuming $R(t=0)=0$)
\begin{align}
I=x_0+\int_{x_0}^xdx'\left(\frac{\sigma}{\gamma(1-x')-\sigma}+1\right)=x+\frac{\sigma}{\gamma}\,\log\left(\frac{1-x}{1-x_0}\right)\,.\label{IformX}
\end{align}
Inserting this into (\ref{FormPreExpgtt2}) we therefore find the metric $g_{tt}^{(2)}$ as a function of $x$ as well as $\gamma,\sigma$ and the initial conditions $x_0$
\begin{align}
g_{tt}^{(2)}=\frac{1-x}{x}\,\left(x\gamma +\sigma\,\,\log\left(\frac{1-x}{1-x_0}\right)\right)^2\,.\label{FormExpgtt2}
\end{align}
We have verified that the dynamics (\ref{IformX}) is indeed compatible with the numerics discussed above and in particular that the form of the metric (\ref{FormExpgtt2}) indeed matches the numerical form plotted in the left panel of Figure~\ref{Fig:SIRshannonUnivnonrec}. Notice in particular, since $x=I+R$ is monotonically growing for the entire dynamics, the metric has only a single branch. Furthermore, the metric has zeroes at
\begin{align}
x=1+\frac{\sigma}{\gamma}\,W\left(-\frac{\gamma}{\sigma}\,(1-x_0)\,e^{-\gamma/\sigma}\right)\,,
\end{align}
which are in fact two values due to the two branches of the Lambert function.
\item metric $g_{tt}^{(3)}$: We can repeat the above analysis for the metric $g_{tt}^{(3)}$ in the third line of (\ref{FormMetricsGenSIR}). Let this time $x(t)=p_1^{(3)}(t)=(S+R)(t)$, such that for $\zeta=0$
\begin{align}
g_{tt}^{(3)}=\frac{(1-x)}{x}\,(\gamma S-\sigma)^2\,.
\end{align}
In order to express $g_{tt}^{(3)}$ entirely in terms of $x$, we need to resolve the dynamics by expressing $S$ as a function of $x$. To this end, we divide the first and third equation in~(\ref{SIRequations})
\begin{align}
&\frac{dR}{dS}=\frac{d(x-S)}{dS}=-\frac{\sigma}{\gamma\,S}&&\text{such that} &&\frac{dx}{dS}=1-\frac{\sigma}{\gamma S}\,.
\end{align}
This differential equation can be integrated, where we consider the initial conditions $S(t=0)=x_0=1-I(t=0)$
\begin{align}
x=x_0+\int_{x_0}^S dS'\,\left(1-\frac{\sigma}{\gamma S'}\right)=S-\frac{\sigma}{\gamma}\,\log(S/x_0)\,.
\end{align}
This relation can be inverted to express $S$ as a function of $x$
\begin{align}
S=-\frac{\sigma}{\gamma}\,W\left(-\frac{x_0 \gamma}{\sigma}\,e^{-\frac{x\gamma}{\sigma}}\right)\,,
\end{align} 
where $W$ is the Lambert function. Notice, however, that this solution is not unique, but has two branches, reflecting the fact that $S+R$ is not a monotonic function throughout the entire time evolution. Indeed, the metric takes the form
\begin{align}
g_{tt}^{(3)}(x)=\frac{1-x}{x}\,\sigma^2\,\left(1+W\left(-\frac{x_0 \gamma}{\sigma}\,e^{-\frac{x\gamma}{\sigma}}\right)\right)^2\,.
\end{align}
and also has two different branches. This is indeed, what we find from the numerical solution of (\ref{SIRequations}), which is plotted in the right panel of Figure~\ref{Fig:SIRshannonUnivnonrec}.
\end{itemize}

\noindent
We only briefly remark that generalising the discussion for the case $\zeta\neq 0$ (for example for the metric $g_{tt}^{(2)}$) is more difficult, due to the fact that the differential equation (\ref{DiffSIRxdep2}) becomes
\begin{align}
\frac{dx}{dI}=1+\frac{\zeta x-(\zeta+\sigma)I}{I(\sigma-\gamma(1-x))}\,,\label{ZetaDefEqI}
\end{align}
which is no longer directly integrable. However, a numerical solutions (see Figure~\ref{Fig:SIRshannonUnivrec}) can nevertheless be obtained, which again reveals a metric that is piecewise of the form (\ref{LVinter}).


\subsection{S$\text{I}^n$ Model}\label{Sect:SInModel}
Combining (and generalising) the ideas of both the Lotka-Volterra model in Section~\ref{Sect:LoktaVolterra} and the SIR model in Section~\ref{Sect:EpidemiologyModels}, we next consider a simple compartmental model with multiple variants of the pathogen. Concretely, we denote $S(t)$ the number of susceptible individuals at time $t\in\mathbb{R}$ and $I_i(t)$ the number of infectious individuals who are infected with the $i$th variant, with $i\in\{1,\ldots,n\}$ (for $n\in\mathbb{N}$ with $n\geq 2$). We assume that each infectious carries exactly one variant and cannot be infected with any of the other variants. Here, we also exclude the possibility of recovery (we shall discuss the generalisation in the next Subsection), such that the dynamics of the epidemiological process is described by the following $n+1$ coupled differential equations
\begin{align}
&\frac{dS}{dt}(t)=-S(t)\,\sum_{i=1}^n\gamma_i\,I_i(t)\,,&&\text{and} &&\frac{dI_i}{dt}(t)=\gamma_i\,S(t)\,I_i(t)\,,\hspace{0.2cm}\forall i=1,\ldots,n\,.\label{SInModel}
\end{align}
Here $\gamma_{1,\ldots,n}\in\mathbb{R}_+$ are constants that describe the infection rate of the $i$th variant. We furthermore consider initial conditions at $t=0$ such that $S(0)+\sum_{i=1}^nI_i(0)=1$ and therefore (since $\frac{d}{dt}\left(S(t)+\sum_{i=1}^nI_i(t)\right)=0$), the size of the population remains constant. For the system (\ref{SInModel}) we can define various different probability distributions. Here we shall be interested in two in particular
\begin{align}
&p^{(1)}:\,\{1,\ldots,n\}\times \mathbb{R}\longrightarrow [0,1]\,,&&\text{with}&&p_i^{(1)}(t):=\frac{I_i(t)}{\sum_{j=1}^n I_j(t)}\,,\nonumber\\
&p^{(2)}:\,\{1,\ldots,n\}\times \mathbb{R}\longrightarrow [0,1]\,,&&\text{with}&&p_i^{(2)}(t):=\frac{\gamma_i\,I_i(t)}{\sum_{j=1}^n\gamma_jI_j(t)}\,,\label{ProbDistributionSII}
\end{align}
corresponding to the (normalised) distribution of (all) infectious individuals or individuals newly becoming infectious at time $t$ ('newly infected') among all variants, respectively. The metrics (\ref{DefFisherInformationMatrix}) associated with these probability distributions can be expressed explicitly in terms of $(S,I_i)$ as follows
\begin{align}
&g_{tt}^{(1)}=\frac{S^2}{\left(\sum_{j=1}^nI_j\right)^2}\sum_{1\leq i<j\leq n}I_i I_j (\gamma_i-\gamma_j)^2\,,&&\text{and} &&g_{tt}^{(2)}=\frac{S^2}{\left(\sum_{j=1}^n\gamma_j I_j\right)^2}\sum_{1\leq i<j\leq n}I_i I_j \gamma_i\gamma_j(\gamma_i-\gamma_j)^2\,,\label{FisherMetricsSIn}
\end{align}

\begin{figure}[htbp]
\begin{center}
\includegraphics[width=7.5cm]{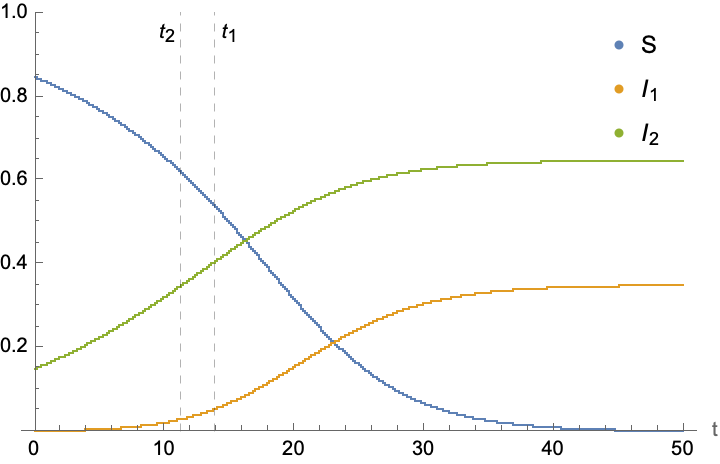}\hspace{1cm}\includegraphics[width=7.5cm]{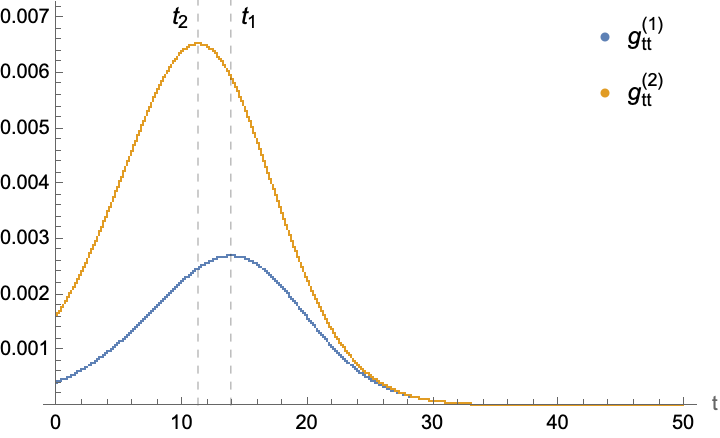}
\end{center}
\caption{Left panel: Time evolution of $(S,I_1,I_2)$ (with $n=2$) as a function of time $t$. The parameters of the model were chosen to be $\gamma_1=0.4$, $\gamma_2=0.1$, $I_1(t=0)=0.001$ and $I_2(t)=0.15$. The plot shows a numerical Runge-Kutta solution of (\ref{SInModel}) with a time-step of $\tau=0.0001$. Right panel: Fisher metrics (\ref{FisherMetricsSIn}) as functions of time. The vertical dashed lines indicate the maxima $t_{1,2}$ of $g_{tt}^{(1,2)}$.}
\label{Fig:SInEpidem}
\end{figure}

\noindent
A typical time evolution of a solution of (\ref{SInModel}) for the simplest example $n=2$ (of multiple variants) is schematically shown in the left panel of Figure~\ref{Fig:SInEpidem} and the metrics (\ref{FisherMetricsSIn}) are plotted in the right panel: here variant 2 was chosen to be more infectious than variant 1, but has a much lower initial condition. Such a setup models the emergence of a new variant (the more infectious variant 2) in a population where variant 1 is already established. One can generalise such a scenario to simulate the emergence of a new variant in a population where already several others are circulating. To this end we shall label the number of infectious $I_i$ in (\ref{SInModel}) in a slightly different fashion: let $\alpha,\beta\in\{2,\ldots,n\}$ with $n\geq 2$ and let
\begin{align}
&\gamma_1\gg \gamma_\alpha\hspace{0.2cm}\text{and} \hspace{0.2cm}|\gamma_\alpha-\gamma_\beta|\ll\,,1\,,&&I_1(t=0)<I_\alpha(t=0)\,,&&\forall \alpha,\beta\in\{2,\ldots,n\}\,,\label{CaseDifferenceVariants}
\end{align}
which simulates the emergence of a more infectious variant (called variant 1) in the presence of an already established class of very similar variants. In this case, the summations in (\ref{FisherMetricsSIn}) can be approximated by
\begin{align}
&g_{tt}^{(1)}\sim\frac{S^2 I_1\bar{I}\, (\gamma_1-\bar{\gamma})^2}{\left(I_1+\bar{I}\right)^2}\,,&&\text{and} &&g_{tt}^{(2)}\sim\frac{S^2\, I_1\bar{I}\,\gamma_1\bar{\gamma} (\gamma_1-\bar{\gamma})^2}{\left(\gamma_1I_1+\bar{\gamma}\bar{I}\right)^2}\,,\label{FisherMetricsSInApprox}
\end{align}
where we have defined
\begin{align}
&\bar{\gamma}=\frac{1}{n-1}\sum_{\alpha=2}^n \gamma_\alpha\,,&&\text{and} &&\bar{I}=\sum_{\alpha=2}^n I_\alpha\,,\label{FormMetGenSII}
\end{align}
which are respectively the average infection rate of the variants $\{I_\alpha\}_{\alpha\in\{2,\ldots,n\}}$ and their combined number of infectious individuals. The metrics (\ref{FisherMetricsSInApprox}), however, are exactly the Fisher metrics for two variants $(I_1,\bar{I})$. Thus, from the perspective of the information in the system, the variants $I_\alpha$ combined behave essentially as a single one (which justifies the use of clustering tools as used for example in \cite{MLvariants}).

We next model the dynamics of the emergence of a new variant using the flow equation (\ref{FlowEquation}) by expressing the metrics $g_{tt}^{(1,2)}$ in(\ref{FormMetGenSII}) as a function of the probability distributions. As explained above, we shall  consider the case $n=2$ in (\ref{SInModel}) as approximation for a new variant (variant 1) emerging among already established ones, which collectively behave as variant 2. We can treat both probability distributions in (\ref{ProbDistributionSII}) in parallel by defining
\begin{align}
&\left\{\begin{array}{lclr}x=\frac{I_1}{I_1+I_2}\,,&\text{such that} &I_2=\frac{1-x}{x}\,I_1\,,&\text{for }p^{(1)}\,,\\[8pt]
y=\frac{\gamma_1 I_1}{\gamma_1 I_1+\gamma_2 I_2}\,,&\text{such that} &I_2=\frac{1-y}{y}\,\frac{\gamma_1}{\gamma_2}\,I_1\,,&\text{for }p^{(2)}\,,\end{array}\right.
\end{align}
and therefore the metric becomes
\begin{align}
&g_{tt}^{(1)}=\frac{1-x}{x}\,(\gamma_1-\gamma_2)^2\,(x-I_1)^2\,,&&g_{tt}^{(2)}=\frac{1-y}{y}\,\frac{(\gamma_1-\gamma_2)^2}{\gamma_2^2}\,\left(y\gamma_2(1-I_1)-\gamma_1 I_1(1-y)\right)^2\,.
\end{align}
To write these metrics entirely as functions of $x$ or $y$, we need to resolve the dynamics of the system to express $I_1$ as a function of $x$ or $y$ respectively. To this end, we divide the two equations (\ref{SInModel})
\begin{align}
&\left\{\begin{array}{lcr}\frac{dI_2}{dI_1}=\frac{1-x}{x}-\frac{I_1}{x^2}\,\frac{dx}{dI_1}=\frac{1-x}{x}\,\frac{\gamma_2}{\gamma_1} & \text{for} & p^{(1)}\,,\\[8pt]
\frac{dI_2}{dI_1}=\frac{\gamma_1}{\gamma_2}\left(\frac{1-y}{y}-\frac{I_1}{y^2}\,\frac{dy}{dI_1}\right)=\frac{1-y}{y} & \text{for}&p^{(2)}\,,
\end{array}\right.
\end{align}
such that
\begin{align}
&\frac{du}{dI_1}=\frac{\gamma_1-\gamma_2}{\gamma_1}\,\frac{u(1-u)}{I_1}\,,&&\text{with} &&u=\left\{\begin{array}{lcl}x & \text{for}& p^{(1)}\,,\\ y & \text{for}& p^{(2)}\,.\end{array}\right.
\end{align}
We can integrate this equation with the initial conditions 
\begin{align}
&I_1(t=0)=I_1^{(0)}\,,&& \text{and} &&u(t=0)=u_0=\left\{\begin{array}{lcl}\frac{I_1}{I_1+I_2}\big|_{t=0} & \text{for}& p^{(1)}\,,\\[8pt] \frac{\gamma_1I_1}{\gamma_1I_1+\gamma_2I_2}\big|_{t=0} & \text{for}& p^{(2)}\,.\end{array}\right.
\end{align}
to find 
\begin{align}
\log(I_1/I_1^{(0)})=\int_{u_0}^u\frac{\gamma_1 dz}{(1-z)z(\gamma_1-\gamma_2)}=\frac{\gamma_1}{\gamma_1-\gamma_2}\,\log\left(\frac{u(1-u_0)}{u_0(1-u)}\right)\,.
\end{align}
which leads to 
\begin{align}
I_1=I_1^{(0)}\,\left(\frac{u(1-u_0)}{u_0(1-u)}\right)^{\frac{\gamma_1}{\gamma_1-\gamma_2}}\,.\label{SIIAnalyticFormI}
\end{align}

\begin{figure}[htbp]
\begin{center}
\includegraphics[width=7.5cm]{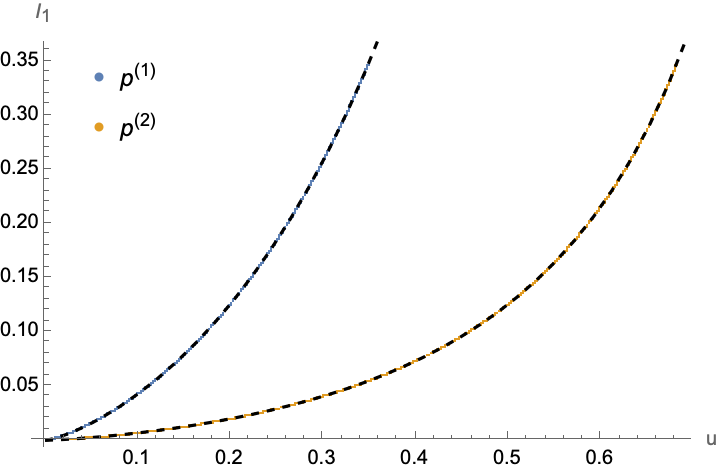}\hspace{1cm}\includegraphics[width=7.5cm]{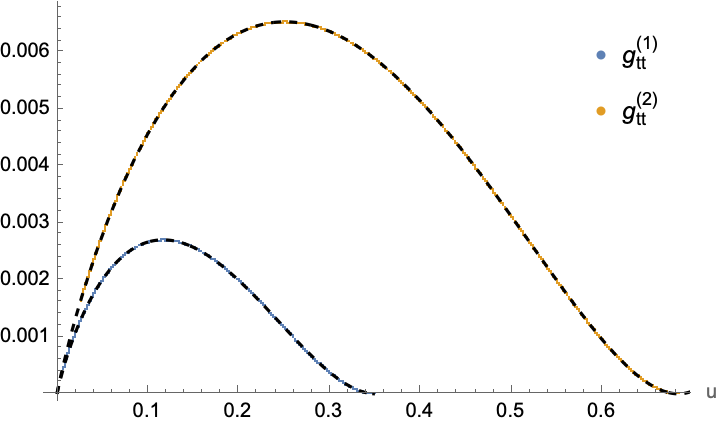}
\end{center}
\caption{Left panel: comparison of the numerical solutions for $I_1$ (coloured lines) with the form (\ref{SIIAnalyticFormI}). Right panel: comparison of the numerical solutions for the metrics $g_{tt}^{(1,2)}$ (coloured lines) with the form (\ref{SIIAnalyticMetric}). Both plots use the same numerical values and initial conditions as in Figure~\ref{Fig:SInEpidem}.}
\label{Fig:SIIAnalyticvsNumerics}
\end{figure}

\noindent
The form of this solution is compared with the numerical solutions discussed above in the left panel of Figure~\ref{Fig:SIIAnalyticvsNumerics}. Furthermore, the metrics become
\begin{align}
g_{tt}^{(1)}&=\frac{1-x}{x}\,(\gamma_1-\gamma_2)^2\,\left(x-I_1^{(0)}\,\left(\frac{x(1-x_0)}{x_0(1-x)}\right)^{\frac{\gamma_1}{\gamma_1-\gamma_2}}\right)^2\,,\nonumber\\
g_{tt}^{(2)}&=\frac{1-y}{y}\,\frac{(\gamma_1-\gamma_2)^2}{\gamma_2^2}\,\left(\gamma_2 y-I_1^{(0)}\,\left(\frac{y(1-y_0)}{y_0(1-y)}\right)^{\frac{\gamma_1}{\gamma_1-\gamma_2}}(y \gamma_2+(1-y)\gamma_1)\right)^2\,,\label{SIIAnalyticMetric}
\end{align}
which are compared to the numerical results in the right panel of Figure~\ref{Fig:SIIAnalyticvsNumerics}. We note that the metrics (\ref{SIIAnalyticMetric}) can be well approximated by a function of the form (\ref{LVinter}) with
\begin{align}
(x_1,x_2,a,b,c)\sim \left\{\begin{array}{lcl}(0,0.347,0.233,0.879,1.745) &\text{for} & g_{tt}^{(1)}\,, \\ (0,0.676,0.087,0.907,1.559) & \text{for} & g_{tt}^{(2)}\,. \end{array}\right.
\end{align}

\subsection{S$\text{I}^n$R(S) Model}\label{Sect:SIIRModel}
As a generalisation of the S$\text{I}^n$ model discussed in the previous Subsection, we now consider an S$\text{I}^n$R(S) model, in which infectious individuals may recover (and potentially may become re-infected at a later time). To this end, in addition to $S$ and $I_i$ (with $i\in 1,\ldots,n$) we introduce as further compartment the number of removed (recovered) individuals $R(t)$ at time $t$. As before, we consider $n\in\mathbb{R}$ variants, with (slightly) different properties regarding infection and removal:
\begin{align}
&\frac{dS}{dt}=-S\,\sum_{i=1}^n\gamma_i\,I_i+\zeta\,R\,,&&\frac{dR}{dt}=\sum_{i=1}^n\sigma_i\,I_i-\zeta\,R\,,&&\frac{dI_i}{dt}=\gamma_i\,S\,I_i-\sigma_i\,I_i\,,\hspace{0.2cm}\forall i=1,\ldots,n\,,\label{SInRModel}
\end{align}
with $\sigma_i,\gamma_i\in\mathbb{R}_+$ denoting the infection and recovery rates of the $i$th variant and $\zeta\in\mathbb{R}_+$ the rate at which removed individuals become susceptible again. We furthermore assume initial conditions at time $t=0$ such that $\left(S+\sum_{i=1}^nI_i+R\right)(t)=1$. For the model (\ref{SInRModel}) we shall consider the following various probability distributions
\begin{align}
&p^{(1)}:\,\{1,\ldots,n\}\times \mathbb{R}\longrightarrow [0,1]\,,&&\text{with}&&p_i^{(1)}(t):=\frac{I_i(t)}{\sum_{j=1}^n I_j(t)}\,,\nonumber\\
&p^{(2)}:\,\{1,\ldots,n\}\times \mathbb{R}\longrightarrow [0,1]\,,&&\text{with}&&p_i^{(2)}(t):=\frac{\gamma_i\,I_i(t)}{\sum_{j=1}^n\gamma_jI_j(t)}\,,\nonumber\\
&p^{(3)}:\,\{1,\ldots,n\}\times \mathbb{R}\longrightarrow [0,1]\,,&&\text{with}&&p_i^{(3)}(t):=\frac{\sigma_i\,I_i(t)}{\sum_{j=1}^n\sigma_jI_j(t)}\,,\label{ProbabilityDistrSIR}
\end{align}
corresponding to the normalised distribution of infectious individuals, individuals becoming infectious ('newly infected') and individuals recovering, respectively. The associated Fisher information metrics (\ref{DefFisherInformationMatrix}) take the form
\begin{align}
g_{tt}^{(1)}&=\frac{\sum_{1\leq i<j\leq n}I_i I_j \left(S(\gamma_i-\gamma_j)-(\sigma_i-\sigma_j)\right)^2}{\left(\sum_{k=1}^n I_k\right)^2}\,,\nonumber\\
g_{tt}^{(2)}&=\frac{\sum_{1\leq i<j\leq n}I_i I_j \gamma_i\gamma_j \left(S(\gamma_i-\gamma_j)-(\sigma_i-\sigma_j)\right)^2}{\left(\sum_{k=1}^n \gamma_kI_k\right)^2}\,,\nonumber\\
g_{tt}^{(3)}&=\frac{\sum_{1\leq i<j\leq n}I_i I_j \sigma_i\sigma_j \left(S(\gamma_i-\gamma_j)-(\sigma_i-\sigma_j)\right)^2}{\left(\sum_{k=1}^n \sigma_kI_k\right)^2}\,.\label{FormMetrics}
\end{align}
These metrics can be written in a unified form as a function of the probabilities $p_{i=1,\ldots,n-1}^{(a)}$
\begin{align}
&g_{tt}^{(a)}=\sum_{i=1}^{n-1}p_i^{(a)}\tau_{in}\left(\tau_{in}-\sum_{j=1}^{n-1} p_j^{(a)}\,\tau_{jn}\right)\,,&&\forall a=1,2,3\,,\label{UniformMetricsSInR}
\end{align}
where we have used $p_n^{(a)}=1-\sum_{i=1}^{n-1}p_i^{(a)}$ for the probabilities (\ref{ProbabilityDistrSIR}). The combinations 
\begin{align}
&\tau_{ij}=S(\gamma_i-\gamma_j)-(\sigma_i-\sigma_j)\,,&&\forall i,j=1,\ldots,n\,,
\end{align}
depend on $S$, such that (\ref{UniformMetricsSInR}) is not a function of the $p_i^{(a)}$ (and thus the $I_{1,\ldots,n}$) alone. Following (\ref{MetEq}), the metric can also be represented in the following factorised form
\begin{align}
&g_{tt}^{(a)}=\sum_{i,j=1}^{n-1}\left(1-\frac{\delta_{ij}}{p_i^{(a)}}\sum_{k\neq i}p_k^{(a)}\right)\,\frac{dp^{(a)}_i}{dt}\,\frac{dp_{j}^{(a)}}{dt}\,,&&\forall a=1,2,3\,.\label{Matrix2d}
\end{align}
Comparing (\ref{Matrix2d}) with (\ref{UniformMetricsSInR}) indeed allows to express the dynamics of the $p_i^{(a)}$ as\footnote{Since (\ref{Matrix2d}) is bilinear in the derivatives of $p_i^{(a)}$, there remains a sign ambiguity which is related to the direction of time.}
\begin{align}
&\frac{d p_i^{(a)}}{dt}=p_i^{(a)}\left(\tau_{in}-\sum_{j=1}^{n-1}\tau_{jn}\,p_j^{(a)}\right)\,,&&\begin{array}{l}\forall a=1,2,3\,,\\\forall i=1,\ldots,n-1\,,\end{array}\label{DynamicsSInR}
\end{align}
which still implicitly depends on $S(t)$. Notice also that the probabilities $p_i^{(a)}$ are not necessarily monotonic functions. For $n=2$, this can be seen by writing the time-derivatives in the following manner\footnote{Here we implicitly assume $\gamma_1\neq \gamma_2$.}
\begin{align}
&\left(S-\frac{\sigma_1-\sigma_2}{\gamma_1-\gamma_2}\right)\,\frac{dp_{1}^{(a)}}{dt}=-\left(S-\frac{\sigma_1-\sigma_2}{\gamma_1-\gamma_2}\right)\,\frac{dp_{2}^{(a)}}{dt}=\frac{g_{tt}^{(a)}}{\gamma_1-\gamma_2}\,,&&\forall a\in\{1,2,3\}\,.\label{DynamicsSInRLong}
\end{align}
Since $S-\frac{\sigma_1-\sigma_2}{\gamma_1-\gamma_2}$ is not positive definite, but $g_{tt}^{(a)}\geq 0$, the sign of the time-derivatives of the probabilities are not fixed.

\begin{figure}[htbp]
\begin{center}
\includegraphics[width=7.5cm]{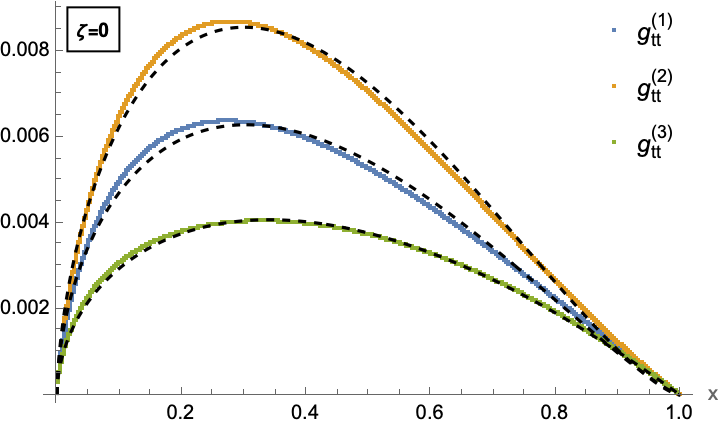}\hspace{1cm}\includegraphics[width=7.5cm]{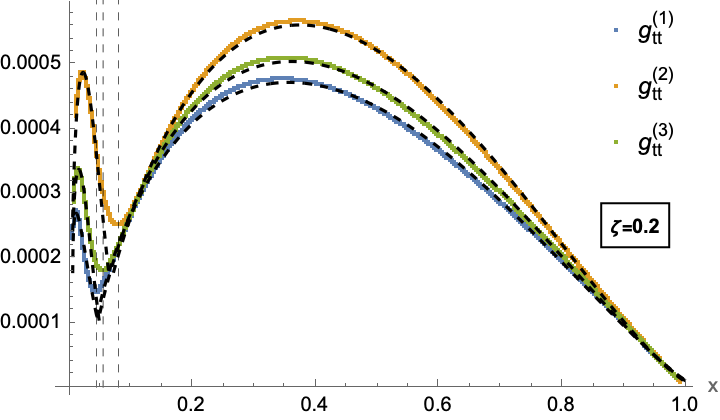}
\end{center}
\caption{Numerical plots of the metrics $g_{tt}^{(1,2,3)}$ as functions of $x=p_1^{(1,2,3)}$ respectively for $\zeta=0$ (left) and $\zeta=0.2$ (right). The dotted black lines give interpolations of the metrics following (\ref{LVinter}), as explained in the text.}
\label{Fig:SInRNumericalMetrics}
\end{figure}

As in the previously discussed examples, in order to write (\ref{DynamicsSInR}) as an effective model for the dynamics of the system, we need to resolve the dynamics. In the present case, we shall do so in the particular case $n=2$ and provide numerical solutions of the metric as functions of $p_1^{(a)}$ in Figure~\ref{Fig:SInRNumericalMetrics}. The dashed black lines give approximations of the metrics following (\ref{LVinter}). Specifically, for $\zeta=0$ (left panel) we find as best fits
\begin{align}
&g_{tt}^{(1)}:&&(x_1,x_2,a,b,c)\sim (0,1,0.019,0.552,1.269)\,,\nonumber\\
&g_{tt}^{(2)}:&&(x_1,x_2,a,b,c)\sim (0,1,0.030,0.612,1.426)\,,\nonumber\\
&g_{tt}^{(3)}:&&(x_1,x_2,a,b,c)\sim (0.011,0.522,1.011)\,,
\end{align}

\noindent
while for $\zeta=0.2$ (right panel) we find the approximation
{\allowdisplaybreaks
\begin{align}
&g_{tt}^{(1)}:&&(x_1,x_2,a,b,c)\sim\left\{\begin{array}{lcl}(0.006,0.120,1.237,0.183,3.375) & \text{for} & \frac{I_1}{I_1+I_2}\in[0,0.044]\,, \\ (0.029,1.023,0.002,0.651,1.324) & \text{for} & \frac{I_1}{I_1+I_2}\in (0.044,1]\,,\end{array}\right.\nonumber\\
&g_{tt}^{(2)}:&&(x_1,x_2,a,b,c)\sim\left\{\begin{array}{lcl}(0.003,0.110,2.825,0.604,2.583) & \text{for} & \frac{\gamma_1I_1}{\gamma_1I_1+\gamma_2I_2}\in[0,0.080]\,, \\ (0.037,1.021,0.002,0.751,1.439) & \text{for} & \frac{\gamma_1I_1}{\gamma_1I_1+\gamma_2I_2}\in (0.080,1]\,,\end{array}\right.\nonumber\\
&g_{tt}^{(3)}:&&(x_1,x_2,a,b,c)\sim\left\{\begin{array}{lcl}(0.001,0.091,34.405,0.656,3.384) & \text{for} & \frac{\sigma_1I_1}{\sigma_1I_1+\sigma_2I_2}\in[0,0.055]\,, \\ (0.033,1.023,0.002,0.682,1.373) & \text{for} & \frac{\sigma_1I_1}{\sigma_1I_1+\sigma_2I_2}\in (0.055,1]\,.\end{array}\right.\nonumber
\end{align}}
We remark that another way to fit these metrics is of the form
\begin{align}
g_{tt}^{(a)}\sim a (x-x_1)^b\,(x_2-x)^c\,\left((x-x_3)(x-\overline{x_3})\right)^{d/2}\,,
\end{align}
with $x_{1,2}$ real- and $x_3\in\mathbb{C}$ a complex zero (and $a,b,c,d$ suitable fitting parameters). Concretely, for the right panel of Figure~\ref{Fig:SInRNumericalMetrics} we find
\begin{align}
&g_{tt}^{(1)}:&&(a,b,c,d,x_1,x_2,x_3)\sim(0.001,0.472,0.825,0.049,0.989,0.908+0.052 i) \,,\nonumber\\
&g_{tt}^{(2)}:&&(a,b,c,d,x_1,x_2,x_3)\sim(0.002,0.1,1.292,0.099,1.011,0.277+0.034 i) \,,\nonumber\\
&g_{tt}^{(3)}:&&(a,b,c,d,x_1,x_2,x_3)\sim(0.002,0.321,1.517,0.063,1.031,-0.466+0.00001 i) \,.
\end{align}

Finally, before closing this Section we remark that for $n=2$,\footnote{Similarly we may assume as in (\ref{CaseDifferenceVariants}) that several variants behave very similarly and can thus be modelled collectively.} we can find exact expressions of $g_{tt}^{(a)}$ as functions of $p_1^{(a)}$ respectively in particular cases: for example we may impose that $S$ and $R$ remain (approximately) constant
\begin{align}
&\frac{dS}{dt}=0=-S(\gamma_1 I_1+\gamma_2 I_2)+\zeta R\,,&&\text{and} &&\frac{dR}{dt}=0=\sigma_1 I_1+\sigma_2 I_2-\zeta R\,.\label{CondStability}
\end{align}
In this case the overall number of infectious individuals remains constant and only the distribution between the two variants changes. The conditions (\ref{CondStability}) are solved by
\begin{align}
&S=\frac{\sigma_1 I_1+\sigma_2 I_2}{\gamma_1 I_1+\gamma_2 I_2}\,,&&\text{and} &&R=\frac{\sigma_1}{\zeta}\,I_1+\frac{\sigma_2}{\zeta}\,I_2\,,
\end{align}
which allows us to simplify the expressions for the metrics:
\begin{align}
g_{tt}^{(1)}(x)&=\frac{x(1-x)\left(\gamma_1\sigma_2-\gamma_2\sigma_1\right)^2}{\left(x(\gamma_1-\gamma_2)+\gamma_2\right)^2}\,,&&\text{with} &&x=\frac{I_1}{I_1+I_2}\,,\nonumber\\
g_{tt}^{(2)}(x)&=\frac{x(1-x)\left(\gamma_1(1-x)+x\gamma_2\right)^2\left(\gamma_1\sigma_2-\gamma_2\sigma_1\right)^2}{\gamma_1^2\gamma_2^2}\,,&&\text{with} &&x=\frac{\gamma_1I_1}{\gamma_1I_1+\gamma_2I_2}\,,\nonumber\\
g_{tt}^{(3)}(x)&=\frac{x(1-x)\left(\sigma_1(1-x)+x\sigma_2\right)^2\left(\gamma_1\sigma_2-\gamma_2\sigma_1\right)^2}{\left((1-x)\gamma_2\sigma_1+x \gamma_1\sigma_2\right)^2}\,,&&\text{with} &&x=\frac{\sigma_1I_1}{\sigma_1I_1+\sigma_2I_2}\,.
\end{align}
In all three cases, the solution of (\ref{FlowEquation}) (with $\alpha=1$) yields the following transcendental equations for $x$ as a function of time
\begin{align}
&p_1^{(1)}:&&t(\gamma_2\sigma_1-\gamma_1\sigma_2)=\gamma_1\,\log(1-x)-\gamma_2\log x\,,\nonumber\\
&p_1^{(2)}:&&t(\gamma_2\sigma_1-\gamma_1\sigma_2)=\gamma_1\,\log(1-x)-\gamma_2\log x-(\gamma_1-\gamma_2)\log\left(\gamma_1(1-x)+\gamma_2x\right)\,,\nonumber\\
&p_1^{(3)}:&&t(\gamma_2\sigma_1-\gamma_1\sigma_2)=\gamma_1\,\log(1-x)-\gamma_2\log x-(\gamma_1-\gamma_2)\log\left(\sigma_1(1-x)+\sigma_2x\right)\,.
\end{align}



\section{General Metric}\label{Sect:GeneralMetric}
In the previous Section, we have re-organised the dynamics of a number of simple examples in the form of the flow equation (\ref{FlowEquation}). To this end, we have defined probability distributions for this models of the form $p:\,\{1,2\}\times \mathbb{R}\to [0,1]$ with $p=(x(t),1-x(t))$ and have expressed the metric $g_{tt}$ as a function of $x$. In all cases, we have seen that the latter can (at least piecewise) be very well approximated by a function of the form (\ref{LVinter}) with $b,c$ constants of order $1$. In this Section we consider the inverse problem, \emph{i.e.} we start with the simplest form of the metric (\ref{LVinter}) and develop the corresponding dynamics in an analytic fashion.
\subsection{Monotonic Evolution}
We start by considering the following simple form of the metric
\begin{align}
&g_{tt}(x)=a\,(x-x_1)(x_2-x)\,,&&\text{for} &&\begin{array}{l}a\in\mathbb{R}_+\,,\\0\leq x_1\leq x\leq x_2\leq 1\,,\end{array}\label{SchematicMetric}
\end{align}
\emph{i.e.} (\ref{LVinter}) for the simplest case $b=c=1$. We assume this metric as the defining input for the dynamics, which fixes $x(t)$ according to the flow equation (\ref{FlowEquation}) for $\alpha=1$. Assuming the initial condition $x(t=t_0)=x_0\in [x_1,x_2]$\footnote{The case $x_1=0$ or $x_2=1$ needs particular care, as we shall discuss below.}, the flow equation can be integrated
\begin{align}
t-t_0=\kappa\,\int_{x_0}^x\frac{dz}{\sqrt{az(1-z)(z-x_1)(x_2-z)}}= \frac{2i\kappa\left(F(u(x),m)-F(u(x_0),m)\right)}{\sqrt{a(1-x_1)x_2}}\,,\label{SolveGenMetric}
\end{align}
where we have assumed that the sign $\kappa\in\{\pm 1\}$ of the first derivative of $x$ remains constant for all times in the interval $[t_0,t]$ and $x_1\leq x\leq x_2$. Furthermore, $F$ in (\ref{SolveGenMetric}) is the incomplete elliptic integral of the first kind 
\begin{align}
F(\phi,k^2)=\int_0^\phi\frac{dz}{\sqrt{1-k^2 \sin^2 z}}\,,
\end{align}
with the arguments
\begin{align}
&m=\frac{x_1(1-x_2)}{x_2(1-x_1)}\,,&&\text{and} &&u(x)=\arcsin\left(\sqrt{\frac{x(1-x_1)}{x_1(1-x)}}\right)\,.
\end{align}
The relation (\ref{SolveGenMetric}) can be solved to express $x$ as a function of time $t$
\begin{align}
x(t)=\frac{x_1\,\text{sn}\left(\tfrac{i\kappa (t-t_0)}{2}\,\sqrt{a(1-x_1)x_2}-F(u(x_0),m),m\right)^2}{1-x_1+x_1\text{sn}\left(\tfrac{i\kappa (t-t_0)}{2}\,\sqrt{a(1-x_1)x_2}-F(u(x_0),m),m\right)^2}\,,\label{SolExpMetr}
\end{align}
where $\text{sn}$ is a Jacobi elliptic function, which can be expressed in terms of Jacobi theta-functions
\begin{align}
&\text{sn}(u,k^2)=\frac{\theta_3(0,\tau)}{\theta_2(0,\tau)}\,\frac{\theta_1(u/\theta_3^2(0,\tau),\tau)}{\theta_4(u/\theta_3^2(0,\tau),\tau)}\,,&&\text{with} &&k=\frac{\theta_2^2(0,\tau)}{\theta_3^2(0,\tau)}\,.
\end{align}
The solution (\ref{SolExpMetr}) is periodic in $t$ with periodicity
\begin{align}
T=\frac{4K(1-m)}{\sqrt{a(1-x_1)x_2}}\,,
\end{align}
where $K$ is the complete elliptic integral of the first kind
\begin{align}
K(m)=F(\pi/2,m)=\frac{\pi}{2}\sum_{n=0}^\infty\left(\frac{(2n-1)!!}{(2n)!!}\right)^2\,m^n\,.
\end{align}

\begin{figure}[htbp]
\begin{center}
\includegraphics[width=7.5cm]{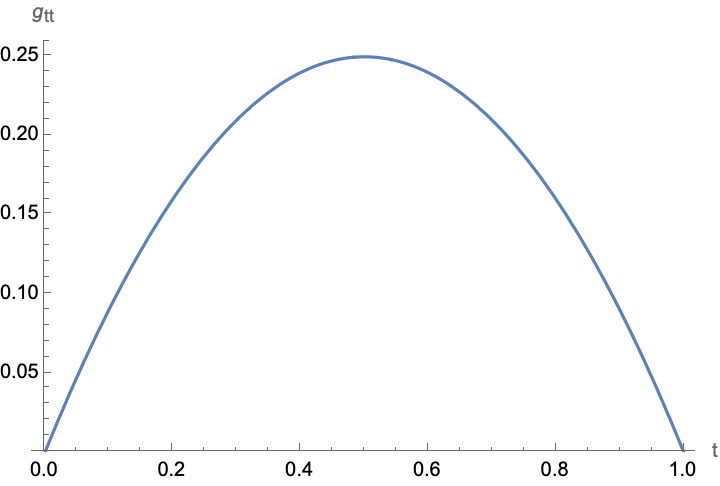}\hspace{1cm}\includegraphics[width=7.5cm]{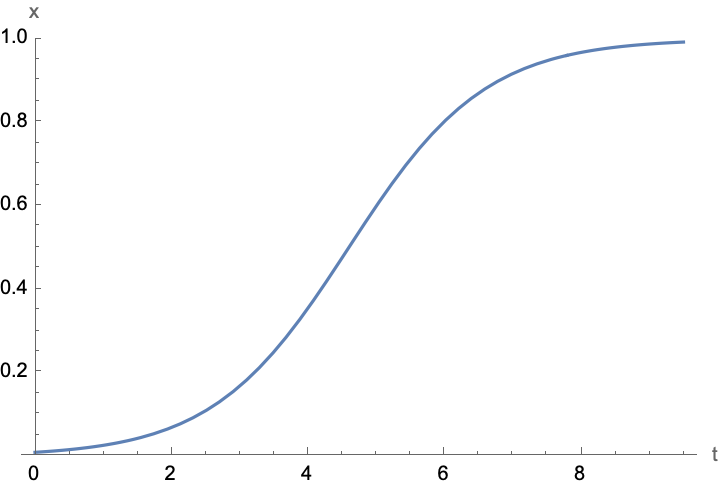}
\end{center}
\caption{Metric (\ref{SchematicMetric}) (left) and $x(t)$ (\ref{SolInterfull}) (right) for $a=1$, $\kappa=1$, $t_0=0$ and $x_0=\tfrac{1}{100}$.}
\label{Fig:GeneralMetricERG}
\end{figure}

In the case of $x_1=0$ and $x_2=1$ care needs to be taken that only $x\in(x_1,x_2)$ is allowed (\emph{i.e.} notably $x_0\notin \{0,1\}$). In this case, the integral in (\ref{SolveGenMetric}) simplifies
\begin{align}
t-t_0=\frac{\kappa}{\sqrt{a}}\,\log\left(\frac{x(1-x_0)}{x_0(1-x)}\right)\,,
\end{align}
such that
\begin{align}
x(t)=\frac{x_0\,e^{\sqrt{a}\,\kappa(t-t_0)}}{1-x_0+x_0\,e^{\sqrt{a}\kappa (t-t_0)}}\,,\label{SolInterfull}
\end{align}
which corresponds to a logistic function and is plotted schematically in Figure~\ref{Fig:GeneralMetricERG}. Notice that 
\begin{align}
\lim_{t\to\infty} x(t)=\left\{\begin{array}{lcl}1&\text{if} & \kappa=1\,,\\0 & \text{if} & \kappa=-1\,,\end{array}\right.
\end{align} 
such that the function $x(t)$ is monotonic for the entire time evolution.
 
\subsection{Full Time Evolution}\label{Sect:FullSchematicTimeEvolution}
Using the result of the previous Subsection, we can construct more complicated time evolutions, by defining the metric $g_{tt}$ to be piecewise of the form (\ref{SchematicMetric}). 
\subsubsection{Flow Between Real Zeroes of the Metric}
Notice that the integral (\ref{SolveGenMetric}) is only defined for $x_1<x<x_2$ (since outside of this interval the integrand becomes imaginary) and for fixed sign of the time derivative $\dot{x}$. In the examples of Section~\ref{Sect:SimpleModels}, we have seen that the zeroes of the metric $x(t)=x_1$ or $x_2$, correspond to points, where the sign of $\dot{x}$ changes, which also leads to a change in the metric. Thus, in this case, we can describe the time evolution by defining the metric piecewise, for fixed sign of $\dot{x}$. In the following we discuss two examples in more detail: the first one inspired by the Lotka-Volterra equation in Section~(\ref{Sect:LoktaVolterra}) and the second one inspired by the SIR model in Section~\ref{Sect:EpidemiologyModels}:

\begin{enumerate}
\item \emph{Periodic Time Evolution:} We first shall consider a metric of the form
\begin{align}
&g_{tt}=\left\{\begin{array}{lcl}a_1\,(x-x_1)(x_2-x) & \text{for} & \dot{x}>0\,, \\ a_2\,(x-x_1)(x_2-x) & \text{for} & \dot{x}<0\end{array}\right.&&\begin{array}{l}a_1,a_2\in\mathbb{R}_+\,,\\0\leq x_1\leq x\leq x_2\leq 1\,,\end{array}\label{MetPeriodic}
\end{align}
where both zeroes of the two branches of the metric coincide (see left panel of Figure~\ref{Fig:GeneralMetricPeriodic}).

\begin{figure}[htbp]
\begin{center}
\includegraphics[width=7.5cm]{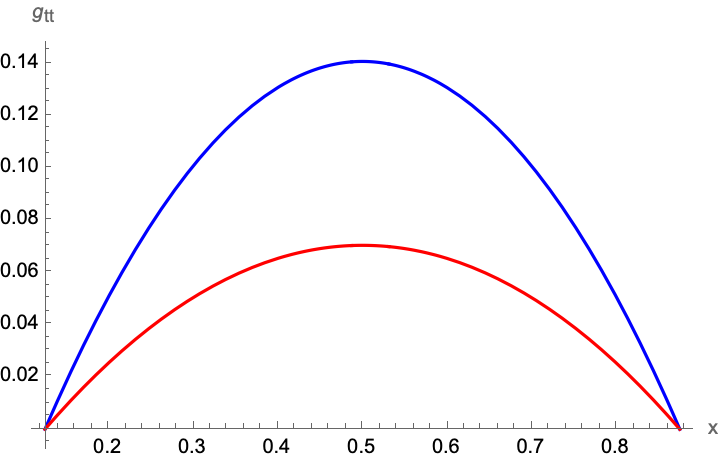}\hspace{1cm}\includegraphics[width=7.5cm]{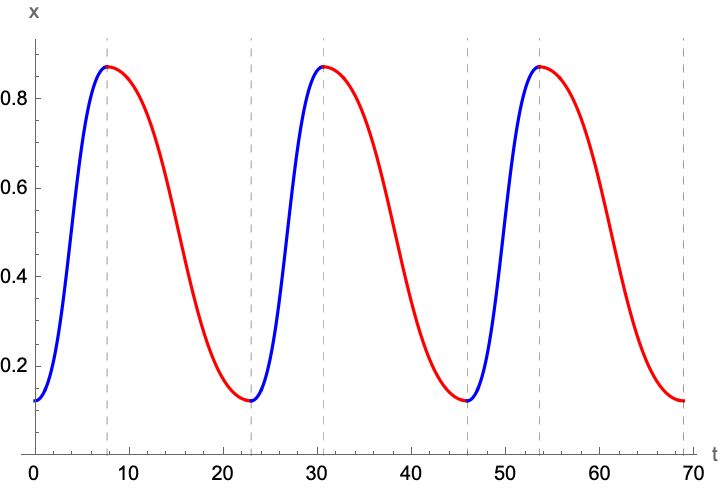}
\end{center}
\caption{Left panel: Metric (\ref{MetPeriodic}) for $x_1=1/8$, $x_2=7/8$, $a_1=1$ (blue curve) and $a_2=1/4$ (red curve). Right panel: corresponding dynamics $x(t)$ (\ref{SolInterfull}) for $t_0=0$ and $x_0=\tfrac{1}{8}$. The dashed lines indicate extrema of $x$, corresponding to zeroes of the metric (\emph{i.e.} points where the branch of the metric changes). The colours of the dynamics $x(t)$ in the right panel match the colours of the branch of the metric in the left panel.}
\label{Fig:GeneralMetricPeriodic}
\end{figure}

\noindent
In this case, the motion of $x(t)$ is periodic and piecewise described by (\ref{SolExpMetr}). The periodicity is given by
\begin{align}
T=\frac{2(\sqrt{a_1}+\sqrt{a_2})\,K\left(\tfrac{x_2-x_1}{(1-x_1)x_2}\right)}{\sqrt{a_1a_2 (1-x_1)x_2}}\,.
\end{align}
as is schematically shown in the right panel of Figure~\ref{Fig:GeneralMetricPeriodic}.

\item \emph{Oscillating Time Evolution:} Let $\{x_1(n)\}_{n\in\mathbb{N}^*}$ and  $\{x_2(n)\}_{n\in\mathbb{N}^*}$ be two (convergent) series of real integers, such that $0<x_1(n)<x_1(n+1)\leq x_2(n+1)<x_2(n)<1$ $\forall n\in\mathbb{N}^*$ and $\lim_{n\to\infty}x_1(n)=\lim_{n\to\infty}x_2(n)=x_f\in[0,1]$. Then consider the series of metrics
\begin{align}
&g_{tt}(n)=\left\{\begin{array}{lcl}a_1\,(x-x_1(n))(x_2(n)-x) & \text{for} & \dot{x}>0\,, \\ a_2\,(x-x_1(n))(x_2(n)-x) & \text{for} & \dot{x}<0\end{array}\right.&&\begin{array}{l}a_1,a_2\in\mathbb{R}_+\,,\\0\leq x_1(n)\leq x\leq x_2(n)\leq 1\,,\end{array}\label{MetPeriodicn}
\end{align}
This series of metrics then describes a time evolution, where $x$ oscillates around $x_f$ and is piecewise described by (\ref{SolExpMetr}). An example for the series
\begin{align}
&x_1(n)=\frac{1}{8}+\frac{9}{4\pi^2}\sum_{m=1}^n\frac{1}{m^2}\,,&&x_2(n)=\frac{7}{8}-\frac{9}{4\pi^2}\sum_{m=1}^n\frac{1}{m^2}\,,&&x_f=\frac{1}{2}\,,\label{DefExSeries}
\end{align}
(and $a_1=a_2=1$) is shown in Figure~\ref{Fig:GeneralMetricDecrease}, leading to a solution oscillating around $x_f$.

\begin{figure}[htbp]
\begin{center}
\includegraphics[width=7.5cm]{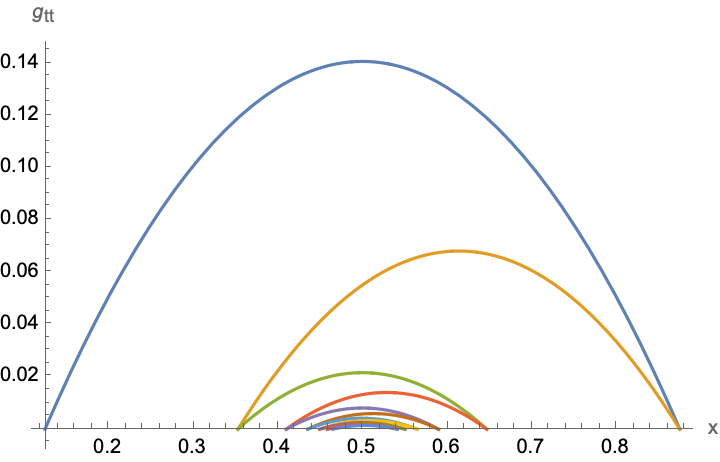}\hspace{1cm}\includegraphics[width=7.5cm]{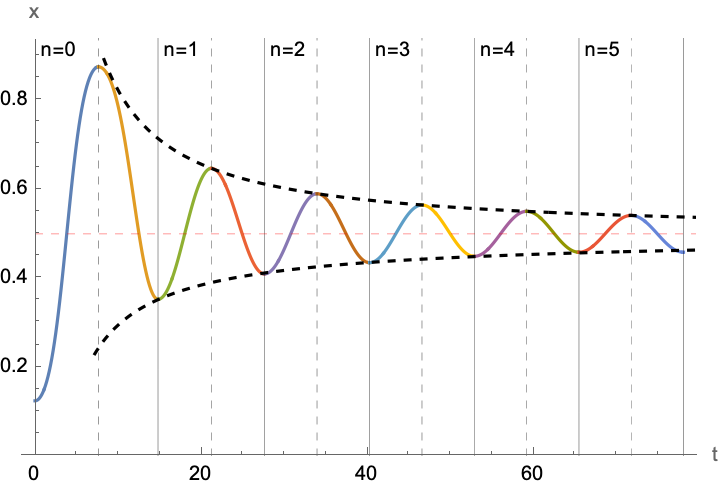}
\end{center}
\caption{Metric (\ref{MetPeriodicn}) (left) and evolution of $x$ (right) using the series (\ref{DefExSeries}) and initial conditions  $t_0=0$ and $x_0=\tfrac{1}{8}$. The black vertical lines indicate extrema of $x$, corresponding to zeroes of the metric: solid lines mark the increase $n\to n+1$, while the dashed lines mark the switch of the branch of the metric in (\ref{MetPeriodicn}). The colours of the dynamics $x(t)$ in the right panel match the colours of the branch of the metric in the left panel.}
\label{Fig:GeneralMetricDecrease}
\end{figure}

\end{enumerate}

\subsubsection{Flow Between Imaginary Zeroes of the Metric}
\begin{figure}[htbp]
\begin{center}
\includegraphics[width=7.5cm]{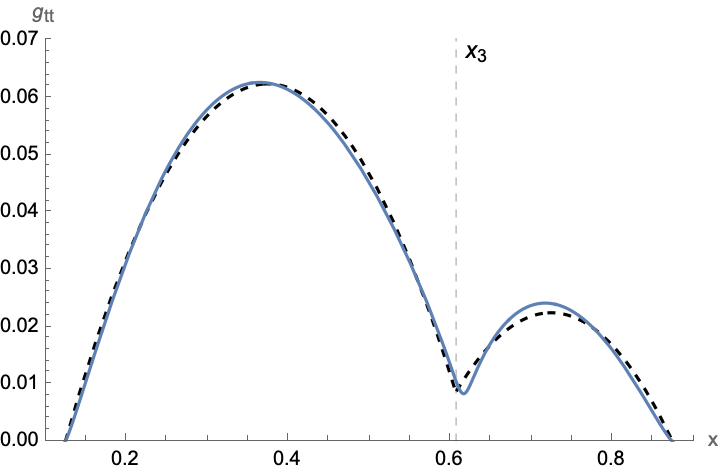}\hspace{1cm}\includegraphics[width=7.5cm]{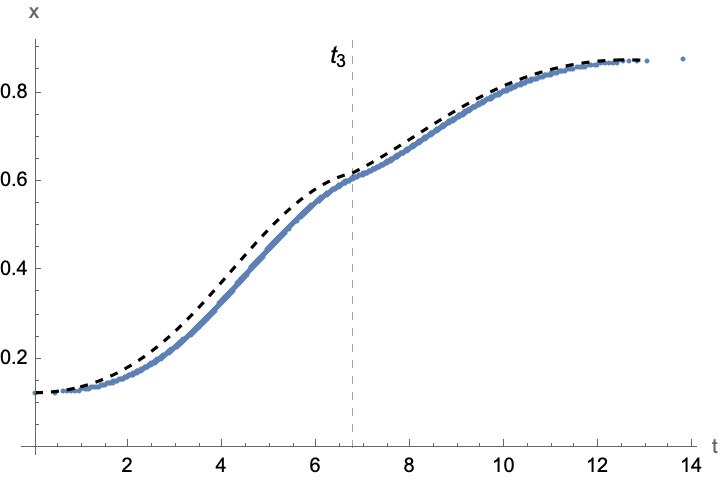}
\end{center}
\caption{Left panel: example of a metric glued along complex zeroes. The dashed black curves are given by $g_{tt}=(x-1/8)(5/8-x)$ for $1/8<x<0.6075$ and $g_{tt}=(x-23/40)(7/8-x)$ for $0.6075<x<1$. Both curves are joined along $x_3=0.6075$ (for which $g_{tt}\neq 0$). The solid blue line represents an approximation of this metric of the form (\ref{ComplexMetricGen}) with $a=1.72$, $b=1.16$, $c=1.24$, $d=0.6$ and $x_1=1/8$, $x_2=7/8$ and $x_3=0.62+0.009 i$. Right panel: corresponding time evolution $x(t)$: the dashed line represents a piecewise solution of the form (\ref{SolExpMetr}), which are joined at $t_3=6.76$. The blue curve is a numerical solution using the approximate metric of the from (\ref{ComplexMetricGen}).}
\label{Fig:ImaginaryMetric}
\end{figure}

Before closing this Section, we want to remark on another possibility for defining metrics that are piecewise of the form (\ref{SchematicMetric}): in the previous examples, we have glued metrics at their (real) zeros, corresponding to extrema of $x$. In principle, we can also glue metrics along points with $g_{tt}\neq 0$, an example of which we have encountered in Section~\ref{Sect:SIIRModel} (see right panel of Figure~\ref{Fig:SInRNumericalMetrics}) in the time evolution of the S$\text{I}^n$R model with re-infection. A simpler example is given by the dashed black curve in the left panel of Figure~\ref{Fig:ImaginaryMetric} (black dashed curve). This metric is approximated by a function of the form
\begin{align}
&g_{tt}=a\, (x-x_1)^b\,(x_2-x)^c\,\left((x-x_3)(x-\overline{x_3})\right)^{d/2}\,,&&\text{with} &&\begin{array}{l}x_1\,,x_2\in[0,1]\text{ with }x_1<x_2\,,\\ x_3\in\mathbb{C}\text{ with }x_1<\text{Re}(x_3)<x_2\,,\end{array}\label{ComplexMetricGen}
\end{align}

\noindent
where $a,b,c,d$ are suitable fitting parameters and $x_{1,2}$ real zeroes of $g_{tt}$, while $x_3$ (and its conjugate $\overline{x_3}$) are complex zeroes. From this perspective, the metric is joined along a complex zero and is not convex on the entire interval $[x_1,x_2]$. The solution of (\ref{FlowEquation}) for the piecewise defined $g_{tt}$ is shown in the right panel of Figure~\ref{Fig:ImaginaryMetric} (black dashed curve), along with a numerical solution using the approximated metric (\ref{ComplexMetricGen}). Indeed, the sign of $dx/dt$ remains positive for the entire time evolution, consistent with the fact that the metric never vanishes. Such a behaviour was used in \cite{cacciapaglia2020evidence} to model periods between two waves of a SARS-CoV-2 epidemic. This behaviour was re-interpreted in \cite{Cacciapaglia:2021cjl} as a system coming close to a(n unstable) fixed point in the context of multiple variants of SARS-CoV-2.

\section{Conclusions}\label{Sect:Conclusions}
In this paper we have studied a reformulation of simple epidemiological- and population dynamical processes in terms of the Fisher information (metric). To this end, we have first re-organised the degrees of freedom of the system into (normalised) probability distributions (in the sense of~(\ref{ProbNorm})) that are functions of time. This is a natural step in epidemiology or population dynamics, since in many approaches a given population is grouped into distinct classes (see \emph{e.g.} compartmental models \cite{Kermack:1927}) and probabilities can be defined as normalised ratios of their sizes. We have associated the Fisher information~\cite{Fisher} with a chosen probability distribution, which (following \cite{Hotelling,Rao,Jeffreys,Lauritzen,amari2000methods}) we have interpreted as a metric $g_{tt}$ of a one-dimensional Riemannian manifold, to highlight its property to capture the distance between the state of the system at different times. In fact, for configurations where the entire probability distribution can be described by a single bounded function $x:\,\mathbb{R}\to [0,\alpha]$ (with $0<\alpha<1$), we have shown that the dynamics can be re-cast in the flow equation (\ref{FlowEquation}), which is entirely determined by the metric. Partially resolving the dynamics of various different models, namely the Lotka-Volterra model (Section~\ref{Sect:LoktaVolterra}), the SIR(S) model (Section~\ref{Sect:EpidemiologyModels}) and compartmental models describing the evolution of two (or more) variants of a disease (Sections~\ref{Sect:SInModel} and \ref{Sect:SIIRModel}), such that $g_{tt}$ can be expressed as a function of $x$ alone, has revealed strong similarities between the different models: the time evolution of $x$ can be described piecewise for regimes in which $x$ is a monotonic function, \emph{i.e.} between two of its local extrema, which correspond to real zeroes of $g_{tt}$. Moreover, in any such regime the metric $g_{tt}$ can be very well approximated simply through powers of its zeroes: if $g_{tt}$ is a convex function it can be approximated through the form (\ref{LVinter}), which can be generalised by including imaginary zeroes (see (\ref{ComplexMetricGen})) if $g_{tt}$ has one or more local minima. The complete dynamics of the system can be described by gluing such metrics together at their zeroes. In fact, using such a metric as a full input for the flow equation (\ref{FlowEquation}) and focusing on the simplest type of metric (\emph{i.e.} (\ref{SchematicMetric})) we have provided an analytic, closed form solution of the monotonic time evolution for $x$ in (\ref{SolExpMetr}). By gluing this solution, we have constructed more general periodic and oscillating solutions in Section~\ref{Sect:FullSchematicTimeEvolution}.

Our approach highlights that the Fisher information metric not only describes families of probabilities, but in fact is capable of capturing their full time evolution provided that enough information about the dynamics of the system can be encoded in it. In our current setting, where the probability distribution is described as a single function $x$, this is tantamount to re-writing $g_{tt}$ as a function of $x$ alone, rather than an explicit function of $t$. The emerging dynamics is quite similar among very different looking models, since details of the system have been resolved, thus revealing universal properties of the underlying time evolution. Specifically, our analysis reveals the importance of zeroes of the metric, both real and imaginary: real zeroes correspond to extrema of the probability distribution $x$ and specifying the metric between two zeroes fixes a monotonic time evolution of $x$. Complex zeroes (or local minima) of $g_{tt}$ provide additional structure (resembling saddle points) to this monotonic time evolution, as can be seen in the right panel of Figure~\ref{Fig:ImaginaryMetric}. 

The importance of (quasi) fixed points in the time evolution was previously pointed out in \cite{cacciapaglia2020evidence,Cacciapaglia:2021cjl} in a concrete epidemiological system: inspired by ideas of renormalisation group methods \cite{Barenblatt,barenblatt_1996,IntermediateAsymptotics,goldenfeld2018lectures,ChenGoldenfeldOono1995,oono1985advances} the (M)eRG (Mutation epidemiological Renormalisation Group) model (based on an original proposal in \cite{DellaMorte:2020wlc,DellaMorte:2020qry}) conjectured a (set of) first order differential equations (called $\beta$-functions or flow equations) that interpolate the cumulative number of individuals infected with (a variant of) a disease between two (disease free) fixed points. This simple model was very successful in describing individual waves of Covid-19 in various regions of the world \cite{cacciapaglia2020second,cacciapaglia2020mining,Cacciapaglia:2020mjf,cacciapaglia2020us,GreenPass} (see also the review \cite{Cacciapaglia:2021vvu}), while the multi-wave structure was attributed to unstable fixed points of the system \cite{cacciapaglia2020multiwave,cacciapaglia2020evidence} which were linked to the appearance of new variants of SARS-CoV-2 replacing previous ones \cite{Cacciapaglia:2021cjl}. Our current analysis is not only in line with this interpretation, but in fact gives a clearer picture of how to describe the notion of 'fixed point' in a more general class of systems: by reducing the dynamics of quite different models (with conceptually different degrees of freedom) to a common form (\ref{FlowEquation}), we can translate the notion of a 'fixed point' (in the language of \cite{cacciapaglia2020evidence,Cacciapaglia:2021cjl}) to (complex) zeroes of the Fisher information, for which moreover a geometric interpretation exists. This allows to re-organise the dynamics of a much larger class of quite different models around these 'fixed points', as is evidenced by our simple approximations in (\ref{LVinter}). Moreover, this re-formulation leads to a potentially simpler description (focused on properties of the system around highly symmetric points), as is showcased by the analytic solutions we have found for the simplest approximation in Section~\ref{Sect:GeneralMetric}.

For the future, we anticipate further applications and generalisations of our work. First of all it will be interesting to extend the description to include cases where the (effective) probability distribution is described by more than one function $x$. Similarly, the inclusion of additional parameters beyond the time-variable will lead to more refined Fisher metrics, capable of describing more sophisticated models and their fixed points. 

\section*{Acknowledgements}
This study is supported by Ministero dell’Università e della Ricerca (Italy), Piano Nazionale di Ripresa e Resilienza, and EU within the Extended Partnership initiative on Emerging Infectious Diseases project number PE00000007 (One Health Basic and Translational Actions Addressing Unmet Needs on Emerging Infectious Diseases). The work of F.S. is
partially supported by the Carlsberg Foundation, semper ardens grant CF22-0922.

\bibliographystyle{ieeetr}
\bibliography{biblio}

\end{document}